\documentclass[aps,prd,reprint,10pt,longbibliography,nofootinbib]{revtex4-2}
\usepackage[utf8]{inputenc}
\usepackage{lmodern}
\usepackage[T1]{fontenc}
\usepackage{graphicx}
\usepackage[
	pdftitle={Point particles in general relativity (and its extensions): beyond linear perturbation theory},
	pdfauthor={Jørgen Musaeus, Adam Pound, Samuel D. Upton},
	pdfdisplaydoctitle,
    pdftex,breaklinks,colorlinks,
    linkcolor=Blue,
    citecolor=teal,
    anchorcolor=red,
    urlcolor=cyan]{hyperref}
\usepackage{mathtools}
\usepackage{amsfonts}
\usepackage{calc}
\usepackage{mathrsfs}
\usepackage{amssymb}
\usepackage{array}
\usepackage[dvipsnames]{xcolor}

\usepackage{bbding}
\renewcommand{\star}{\text{\tiny \FiveStarOpen}}
\usepackage{orcidlink}

\newcommand{\beq}{\begin{equation}}
\newcommand{\eeq}{\end{equation}}

\newcommand{\e}{\varepsilon}
\newcommand{\nhat}{\hat{n}}

\newcommand{\deltaG}{\delta G}
\renewcommand{\SS}{\mathrm{SS}}

\newcommand{\calE}{\mathcal{E}}
\newcommand{\barh}{\bar{h}}

\newcommand{\calP}{\mathcal{P}}
\newcommand{\calR}{\mathcal{R}}
\renewcommand{\S}{\mathrm{S}}
\newcommand{\R}{\mathrm{R}}

\newcommand{\Teff}{\tilde{T}}

\newcommand{\<}{\left\langle}
\renewcommand{\>}{\right\rangle}

\DeclareMathOperator*{\FP}{\mathrm{FP}}

\renewcommand{\O}{\mathcal{O}}

\begin{document}

\title{Point particles in general relativity:\\beyond linear perturbation theory}
\author{Jørgen Musaeus\,\orcidlink{0000-0003-4869-0293}}
\author{Adam Pound\,\orcidlink{0000-0001-9446-0638}}
\author{Samuel D.\ Upton\,\orcidlink{0000-0003-2965-7674}}
\affiliation{School of Mathematical Sciences and STAG Research Centre, University of Southampton, Southampton, United Kingdom, SO17 1BJ}

\date{\today}

\begin{abstract}
Point particles are generically ill defined in fully nonlinear general relativity, although they are naturally well defined in linear perturbation theory. In a previous paper, two of us showed how they can be rigorously defined from matched asymptotic expansions at second order in perturbation theory on a generic background spacetime. Here we show how the resulting field equations can be recast in a more practical, more fully ``skeletonized'' form. We also show that this form of the field equations enables simple derivations of equations of motion. Finally, we show how calculations in second-order perturbation theory can bypass the use of singular ``punctures'' and make rigorous use of off-the-shelf regularization methods such as Blanchet and Damour's Hadamard regularization, opening new avenues to second-order self-force calculations in black hole binaries.
\end{abstract}

\maketitle

\tableofcontents

\section{Introduction}

Point particles are a mainstay of classical physics. They are a core tool in how we learn and apply the principles of Newtonian mechanics and classical electromagnetism. However, they have a thorny history in general relativity (GR). On one hand, a classic result due to Geroch and Traschen~\cite{Geroch:1986jjl} established that point particles are generically ill defined in fully nonlinear GR, in the sense that a point particle is too singular to source a metric whose curvature is well defined as a distribution. On the other hand, point particles are commonly used in approximation schemes within GR, typically combined with regularization schemes. Our goals in this paper are to elaborate on that dichotomy by elucidating the interpretation of point particles in GR and developing practical forms of perturbative field equations with point-particle sources---specifically, field equations at second perturbative order on a generic background spacetime.

\subsection{Perspectives on point particles}

The basic spirit of point-particle treatments in GR stems from Mathisson's ``gravitational skeleton''~\cite{Mathisson:1937zz,Mathisson:2010opl,Dixon:2015vxa}. If we consider a small and light body immersed in a background metric $g_{\alpha\beta}$, then the body creates a small perturbation $h_{\alpha\beta}$. If $g_{\alpha\beta}$ is a vacuum metric, then the linearized Einstein equation for the perturbation reads
\begin{equation}\label{eq:linearized EFE}
    \delta G^{\alpha\beta}[h] = 8\pi T^{\alpha\beta},
\end{equation}
where $T^{\alpha\beta}$ represents a leading-order approximation to the stress-energy tensor of the small body (i.e., neglecting self-interaction). Mathisson showed that the stress-energy tensor can always be decomposed into a sum of Dirac-delta terms,
\begin{multline}\label{eq:T skeleton}
    T^{\alpha\beta} = \int_\gamma \Bigl\{m u^\alpha u^\beta \delta^4(x,x_p) + \nabla_\gamma\bigl[u^{(\alpha} S^{\beta)\gamma} \delta^4(x,x_p)\bigr] \\
    + \nabla_\gamma\left[u^{\alpha} u^{\beta}M^{\gamma} \delta^4(x,x_p)\right] + \ldots\Bigr\}d\tau,
\end{multline}
where $\delta^4(x,x_p)=\delta^4(x^\alpha-x^\alpha_p)/\sqrt{-g}$ is the covariant delta distribution~\cite{Poisson:2011nh}, $\nabla_\alpha$ is the covariant derivative compatible with $g_{\alpha\beta}$, $\gamma$ is the body's representative worldline, with coordinates $x^\alpha_p$ and four-velocity $u^\alpha=dx^\alpha_p/d\tau$, and $\tau$ is proper time. The coefficients in this sum are the body's multipole moments: $m$, its mass; $S^{\alpha\beta}$, its spin dipole moment; $M^\alpha$, its mass dipole moment (which can be set to zero to enforce that $\gamma$ is the body's center of mass); and so on to higher multipole order. This point-particle singularity living on $\gamma$, equipped with the body's set of multipole moments, is the body's ``skeleton''.\footnote{Mathisson actually presented this framework, essentially in the language of distribution theory, \emph{without} utilizing Dirac deltas. It was then recast in terms of $\delta$ functions by Tulczyjew~\cite{Tulczyjew:1959}. See also Ref.~\cite{Steinhoff:2012rw}, for example.} Conservation of stress-energy, $\nabla_\beta T^{\alpha\beta}=0$, then determines evolution equations for $u^\alpha$ and $S^{\alpha\beta}$ in terms of the body's higher moments.

Via the Einstein equation~\eqref{eq:linearized EFE}, the multipole moments in $T^{\alpha\beta}$ directly characterize the behaviour of the metric perturbation near the particle. In Cartesian Fermi-Walker coordinates $(\tau,x^i)$ centered on $\gamma$~\cite{Poisson:2011nh}, the multipole moments enter the perturbation near the particle in the schematic form~\cite{Thorne:1984mz,Pound:2012dk}
\begin{subequations}\label{eq:multipole moments in h}
\begin{align}
    h_{\tau\tau} &\sim \frac{m}{\rho}\bigl[1+{\cal O}(\rho^2)\bigr]  + \frac{M_i n^i}{\rho^2}\bigl[1+{\cal O}(\rho^2)\bigr]\nonumber\\
    &\quad + \frac{M_{ij} n^i n^j}{\rho^3}\bigl[1+{\cal O}(\rho^2)\bigr] + \ldots,  \\
    h_{\tau i} &\sim \frac{S_{ij} n^j}{\rho^2}\bigl[1+{\cal O}(\rho^2)\bigr] + \frac{S_{ijk} n^jn^k}{\rho^3}\bigl[1+{\cal O}(\rho^2)\bigr] \nonumber\\
    &\quad + \ldots,
\end{align}
\end{subequations}
where $\rho\coloneqq\sqrt{\delta_{ij}x^i x^j}$ is proper (orthogonal) distance from~$\gamma$, $n^i\coloneqq x^i/\rho$ is a radial unit vector, and indices are lowered with $\delta_{ij}$. Here, to illustrate the structure, we have included the mass and spin quadrupole moments, $M_{ij}$ and $S_{ijk}$, in addition to the monopole and dipole from Eq.~\eqref{eq:T skeleton}. 
We have not written $h_{ij}$ as it does not feature any additional multipole moments. 
Equation~\eqref{eq:multipole moments in h} is a small-$\rho$ expansion, but with infinitely many negative powers of $\rho$; the $\ell$th moments first appear at order $1/\rho^{\ell+1}$, with the order-$\rho^2$ corrections to each term arising from the tidal moments of the background metric $g_{\alpha\beta}$.

Since Eq.~\eqref{eq:linearized EFE} is linear, all of the above is straightforwardly well defined in classical distribution theory. But at nonlinear orders, when the field interacts with itself, classical distribution theory becomes delicate because products of distributions are ill defined. There are then several viable ways to understand the point-particle limit, either starting from the full stress-energy tensor of an extended material body; from a point-particle skeleton ansatz; or from the field \emph{outside} the body.

First consider the approach of working with an extended material body (hence excluding black holes). If the exact nonlinear problem is reformulated in terms of effective fields and effective multipole moments that remain well defined when the body is shrunk toward zero mass and size, then the point-particle limit can be safely taken. This approach is exemplified by the work of Harte~\cite{Harte:2011ku,Harte:2014wya,Harte:2025tmd}, following the seminal work of Dixon~\cite{Dixon:1964cjb,Dixon:1970zz,Dixon:1970zz,Dixon:1974xoz} (see also Refs.~\cite{Geroch:1975uq,Ehlers:2003tv,Geroch:2017hdb}). In Dixon's formalism, the body's multipole moments are defined as integrals over the body's interior, and one is able to obtain exact, fully nonlinear laws of motion and precession. These laws, which exactly encode the entire content of stress-energy conservation, dictate the evolution of the body's linear and angular momenta in terms of forces and torques arising from the body's interaction with the spacetime metric. 

Dixon's evolution equations, though exact, can only be written in terms of the body's multipole moments (its skeleton) \emph{if} one treats the body as a \emph{test} body that does not correct the spacetime, meaning the metric in the exact equations becomes an external background rather than the full physical metric. In that case, Dixon's laws of motion and precession reduce to the same equations derived by Mathisson (but extended to all multipole orders), which can be obtained from conservation of the stress-energy skeleton~\eqref{eq:T skeleton}.  

Harte and collaborators extended Dixon's results by showing that there exists a transformation from the full physical metric, ${\sf g}_{\alpha\beta}$, to an \emph{effective} metric, $\tilde g_{\alpha\beta}$, that removes the contribution from the body's ``self-field''. This effective metric can be defined such that the body's exact laws of motion and precession take the form of Dixon's test-body equations in $\tilde g_{\alpha\beta}$, expressed in terms of redefined, effective higher multipole moments. For an appropriate transformation $g_{\alpha\beta}\to \tilde g_{\alpha\beta}$~\cite{Harte:2025tmd}, the effective metric $\tilde g_{\alpha\beta}$ and the laws of motion and precession remain well behaved in the point-particle limit. One can then construct an effective stress-energy skeleton, $\tilde T^{\alpha\beta}$, which takes the form of the test-body skeleton~\eqref{eq:T skeleton} but with the background $g_{\alpha\beta}$ replaced by the effective metric $\tilde g_{\alpha\beta}$~\cite{Rahman:2026qho}. The laws of motion and precession are then equivalent to conservation in the effective metric, 
\begin{equation}\label{eq:TDet conservation}
\tilde\nabla_\beta \tilde T^{\alpha\beta}=0.    
\end{equation}

While elegant, this approach has limitations. It is only valid for material bodies rather than black holes, and it does not (yet) provide a clear algorithm for finding the effective metric at nonlinear orders. Nor does it provide a practical field equation or technique for computing the metric perturbations produced by small objects.

A second approach is a bottom-up ``point-particle effective field theory'' (EFT)~\cite{Goldberger:2004jt,Galley:2008ih, Zimmerman:2015rga,Porto:2016pyg,Cheung:2023lnj,Modrekiladze:2026twz}, in which one assumes that, on scales much larger than the object, the object's contribution to the Lagrangian can be replaced with a point-particle contribution with any desired multipole structure. Varying the Lagrangian with respect to the full spacetime metric formally recovers the skeleton~\eqref{eq:T skeleton}, but now in terms of the full metric rather than a background metric or effective metric. Such an expression would be purely formal because it involves the full metric evaluated at the particle's position, where it is singular. In the point-particle EFT, such singularities are attributed to having neglected small-scale physics, and a regularization scheme is adopted to deal with them. Missing information about the small-scale physics (such as the dynamics of the object's quadrupole and higher moments) is then obtained through calculations of physical processes that probe the scale of the object.

The point-particle EFT approach is powerful, applies to both material bodies and black holes, and enables streamlined calculations. However, its application generally relies on methods inherited from high-energy physics, which are often conceptually foreign to classical relativists. Its major successes have also been in the special case of perturbations of flat spacetime, rather than the general case of a small body immersed in a generic curved spacetime. Moreover, it does not directly address Geroch and Traschen's result~\cite{Geroch:1986jjl}, as it does not (to our knowledge) clarify how and whether a point particle can be incorporated as a source in the Einstein field equations in the usual sense of partial differential equations (PDEs). Even in the traditional schools of post-Newtonian theory, which avoid much of the language of high-energy physics, the use of a point-particle ansatz with a regularization scheme~\cite{Blanchet:2013haa} can raise questions about how such regularization relates to the full physical problem of an extended body in nonlinear GR. 

A third approach is based on matched asymptotic expansions, which likewise applies for both material bodies and black holes. This approach, initiated by D'Eath and others in the 1970s and '80s~\cite{DEath:1975jps, Kates:1980ks,Kates:1980zz, Damour:1982wm, Thorne:1984mz,Futamase:1985kcc}, is top-down rather than bottom-up. It proceeds from analysis of the field equations \emph{outside} the body, from which equations like~\eqref{eq:multipole moments in h} can be derived without specification of a stress-energy tensor. 
This is the approach we adopt here. However, one of our principal aims is to show that it implies skeletonized field equations from which one can establish the validity of ``off the shelf'' regularization methods, which opens up new computational avenues and takes a step toward unifying matched expansions with the point-particle EFT approach. 

Our focus will specifically be on gravitational self-force theory, where the method of matched expansions has been pursued the furthest and been most central to deriving foundational results~\cite{Mino:1996nk,Mino:1997wh,Mino:1997bw,Detweiler:2000gt,Poisson:2003wz,Detweiler:2005kq,Rosenthal:2006nh,Rosenthal:2006iy,Fukumoto:2006gv,Gralla:2008fg,Pound:2009sm,Pound:2010pj,Gralla:2010cd,Detweiler:2011tt,Pound:2012nt,Gralla:2012db,Pound:2012dk,Gralla:2013rwa,Pound:2017psq}.\footnote{We refer to Refs.~\cite{Itoh:1999bf,Itoh:2001np,Itoh:2003fy,Futamase:2007zz,Itoh:2009rz} for a matched-expansions approach specialized to the post-Newtonian context. But we note that, in principle, the methods we describe here also apply in the post-Newtonian limit, simply treating the external spacetime as a post-Newtonian expansion.} Self-force theory is often described as a particular approach to the two-body problem in the case of a small body orbiting a much larger one, but it is more broadly a formalism for treating small bodies interacting with external spacetimes. Reference~\cite{Barack:2018yvs} provides a pedagogical introduction; Ref.~\cite{Pound:2015tma}, a more advanced one. The idea of matched expansions in this context is to use multiple asymptotic expansions, one over the exterior spacetime, on scales much larger than the body, and one that zooms in the body. Concretely, suppose the body has mass $m$ and the external spacetime's smallest length scale is $L\gg m$. Defining $\e\coloneqq m/L$, we assume that on the scale $L$, we can expand the full spacetime metric as
\begin{equation}
   {\sf g}_{\alpha\beta} = g_{\alpha\beta} + \e h^{(1)}_{\alpha\beta} + \e^2 h^{(2)}_{\alpha\beta} + \O(\e^3).
\end{equation}
This \emph{outer} expansion treats the small body as the source of a small perturbation to the external universe. Sufficiently close to the body, at distances $\rho\sim m$, this expansion breaks down, as the metric becomes dominated by the body's own gravity. We then assume the existence of a second, \emph{inner} expansion that zooms in on this region: defining scaled Fermi-Walker coordinates, $X^\alpha\coloneqq(\tau,x^i/\e)$, we expand ${\sf g}_{\alpha\beta}$ in powers of $\e$ while holding $X^\alpha$ fixed, such that $\rho$ is of order $\e$. Finally, we assume a matching condition: when the outer expansion is re-expanded for small distances $\rho\ll L$ and the inner expansion is re-expanded for small $\e$ at fixed $\rho$ (or equivalently, for distances $\rho\gg m=\e L$), the two resulting double expansions must agree order by order in $\e$ and $\rho$. 

Together, these three assumptions effectively state there are only two scales in the problem, and that the two expansions tailored to the two scales commute with one another. Recently, Hintz has proved that one-parameter families of solutions to the Einstein equations do exist that satisfy a particularly strong flavor of these conditions, thereby justifying the assumptions of matched expansions~\cite{Hintz:2023lyt,Hintz:2024uwc,Hintz:2024dhy}, at least in the case of small black holes. 

The core outputs of the matched-expansions approach are local expressions for the metric perturbations $h^{(n)}_{\alpha\beta}$ near $\gamma$, as determined by the matching condition together with a local expansion of the Einstein equations. This generalizes the linear result~\eqref{eq:multipole moments in h}, order by order in $\e$. As we alluded to above, the multipole moments in this approach are defined from the metric outside the body, rather than from integrals over the body's interior. Moreover, the field equations \emph{outside} the body also determine the same equations of motion and precession as one would extract from stress-energy conservation---a long-known if potentially surprising fact~\cite{Weyl:1921,Eddington:1924,Einstein:1938yz,Einstein:1949,Infeld:1949}. 

If the local metric obtained outside the body is analytically continued down to $\rho=0$, then it diverges on $\gamma$. More explicitly, the local matched-expansions analysis neatly divides the metric perturbations into singular ``self-fields'', which are specified by the body's multipole moments, and regular ``external fields'' that are specified by external boundary conditions~\cite{Pound:2012dk},
\begin{equation}
 h^{(n)}_{\alpha\beta} = h^{\S(n)}_{\alpha\beta} + h^{\R(n)}_{\alpha\beta}.
\end{equation}
The sum of the external background and the regular fields defines a smooth effective metric,
\begin{equation}
 \tilde g_{\alpha\beta} = g_{\alpha\beta} + \sum_{n\geq1}\e^n h^{\R(n)}_{\alpha\beta},
\end{equation}
which satisfies the vacuum Einstein equation even at $\rho=0$. Although it is not known how this effective metric relates to Harte's beyond linear order, the body \emph{is} known to move as a test body in it (at least at low orders in $\e$~\cite{Detweiler:2000gt,Poisson:2011nh,Gralla:2008fg,Pound:2009sm,Pound:2012nt,Pound:2017psq,Rahman:2026qho}), and $h^{\R(1)}_{\alpha\beta}$ is the Detweiler-Whiting regular field~\cite{Detweiler:2002mi,Poisson:2011nh} in both the matched-expansions and Harte approach. 

In this way, rather than directly providing or assuming a skeletonized stress-energy tensor, matched expansions provides a ``puncture'': a singularity in the spacetime metric, representing the body's self-field, equipped with the body's multipole moments, and moving as a test body on a representative worldline $\gamma$ in an effective metric. 

As we briefly review below, this puncture representation of a point particle does provide a practical scheme for finding $h^{(n)}_{\alpha\beta}$ everywhere, not only near $\gamma$, and to all orders $\e^n$~\cite{Pound:2012dk}. In fact, it provides the only method that has been used in practice at second order~\cite{Miller:2023ers,Upton:2025bja,Leather:2026zhl}. However, there are both conceptual and pragmatic advantages in formulating the field equations in terms of a skeletonized stress-energy tensor. In this paper we aim to formulate such field equations and demonstrate their advantages at second perturbative order.

\subsection{This paper}

Our derivations in the body of the paper are highly technical, which could obscure our overarching narrative. In the remainder of this Introduction, we provide a high-level overview of the main concepts and results. Here and in what follows, rather than using $\e=m/L$, we treat $\e$ as a counting parameter, set equal to unity at the end of a calculation, in order to keep factors of $m$ explicit in~$h^{(n)}_{\alpha\beta}$. 

At first order in perturbation theory, the metric perturbation in the outer expansion, $h^{(1)}_{\alpha\beta}$, satisfies the vacuum Einstein equation at every point away from the representative worldline $\gamma$,
\begin{equation}\label{eq:EFE1 off gamma}
    \delta G^{\alpha\beta}[h^{(1)}] = 0 \quad \text{for }x\notin\gamma.
\end{equation}
The outcome of the matched-expansions analysis is that  the solution to this equation must behave as~\cite{Pound:2012dk}
\begin{equation}\label{eq:h1 near gamma}
    h^{(1)}_{\alpha\beta} \sim \frac{m}{\rho} 
\end{equation}
near $\gamma$. Unlike in Eq.~\eqref{eq:multipole moments in h}, only the monopole appears here because of our assumption that $m$ provides the body's only scale, which forces its multipole moments to scale as $\e^{\ell+1}$. Each successive order in perturbation theory then introduces one additional pair of (mass and spin) multipole moments~\cite{Pound:2012dk,Rahman:2026qho}.

One could imagine directly solving Eq.~\eqref{eq:EFE1 off gamma} subject to the boundary condition~\eqref{eq:h1 near gamma}~\cite{Pound:2010pj}. Equivalently, we can use a puncture scheme~\cite{Barack:2018yvs,Pound:2012dk}. First construct the local self-field $h^{\S(1)}_{\alpha\beta}$ as an explicit small-$\rho$ expansion~\cite{Pound:2012dk,Pound:2014xva}. Define the \emph{puncture} $h^{\calP(1)}_{\alpha\beta}$ as the truncation of that expansion at order $\rho$ or higher, tapered down to zero at some distance from $\gamma$ (either sharply or smoothly). In concert, define the \emph{residual field} $h^{\calR(1)}_{\alpha\beta}\coloneqq h^{(1)}_{\alpha\beta}-h^{\calP(1)}_{\alpha\beta}$, such that $h^{\calR(1)}_{\alpha\beta}$ reduces to $h^{\R(1)}_{\alpha\beta}$ on $\gamma$ but reduces to $h^{(1)}_{\alpha\beta}$ far away from $\gamma$. Moving the puncture to the right-hand side of Eq.~\eqref{eq:EFE1 off gamma}, we obtain a field equation for the residual field,
\begin{equation}\label{eq:EFE1 eff}
    \delta G^{\alpha\beta}[h^{\calR(1)}] = - \delta G^{\alpha\beta}[h^{\calP(1)}]^\star,
\end{equation}
with an effective source. Here we adopt the $\star$ notation from Ref.~\cite{Upton:2021oxf}, which denotes evaluation as an ordinary pointwise function at points off $\gamma$ and extension to $\gamma$ by continuity (when possible). The simplest example is the action of the flat-space Laplacian on a Coulomb potential: $\nabla^2(1/\rho)$ is a $\delta$ function when treated as a distribution, but $\nabla^2(1/\rho)^\star=0$. If we impose continuity of $h^{\calR(1)}_{\alpha\beta}$ at $\gamma$, then the total field $h^{\calR(1)}_{\alpha\beta}+h^{\calP(1)}_{\alpha\beta}$ obtained by solving Eq.~\eqref{eq:EFE1 eff} (and adding on the puncture) is identical to the $h^{(1)}_{\alpha\beta}$ obtained by solving Eq.~\eqref{eq:EFE1 off gamma} subject to \eqref{eq:h1 near gamma}.

One could stop at this puncture scheme. But we can go further by linking the outcome of matched expansions to a traditional point-particle source. If we take the local form of $h^{(1)}_{\alpha\beta}$ obtained from matched expansions and treat $\delta G^{\alpha\beta}[h^{(1)}]$ as a distribution (i.e., as a linear operator acting on an integrable function), then we can \emph{define} the first-order effective stress-energy tensor as 
\begin{equation}\label{eq:EFE1 distributional}
    \delta G^{\alpha\beta}[h^{(1)}]\coloneqq 8\pi T_{(1)}^{\alpha\beta}.
\end{equation}
A brief distributional analysis establishes~\cite{Gralla:2008fg}
\begin{equation}
    T_{(1)}^{\alpha\beta} = m \int_\gamma u^\alpha u^\beta \delta^4(x,x_p)d\tau,
\end{equation}
which is the familiar monopole term from Eq.~\eqref{eq:T skeleton}. Once one knows Eq.~\eqref{eq:EFE1 distributional} is valid, one can start from it and solve directly for $h^{(1)}_{\alpha\beta}$ rather than employing the puncture scheme. 

Now consider the same steps at second order. The second-order perturbation $h^{(2)}_{\alpha\beta}$ satisfies the second-order vacuum Einstein equation at points off $\gamma$,
\begin{equation}\label{eq:EFE2 off gamma}
    \delta G^{\alpha\beta}[h^{(2)}] = -\delta^2 G^{\alpha\beta}[h^{(1)},h^{(1)}] \quad \text{for }x\notin\gamma,
\end{equation}
where $\delta^2 G^{\alpha\beta}$ comprises the quadratic terms in the expansion of the Einstein tensor on the background $g_{\alpha\beta}$, given in Eq.~(A3) of Ref.~\cite{Upton:2021oxf}. In this paper, we consistently work in the Lorenz gauge, in which case the local solution to Eq.~\eqref{eq:EFE2 off gamma}, as determined from matched expansions, has the form~\cite{Pound:2014xva}
\begin{equation}\label{eq:h2 near gamma}
    h^{(2)}_{\alpha\beta} \sim \frac{m^2+M_i n^i}{\rho^2} + \frac{m h^{R(1)}}{\rho} +\ldots 
\end{equation}
near $\gamma$.\footnote{In the special case of the ``highly regular gauge'', the $1/\rho^2$ term vanishes in $h^{(2)}_{\alpha\beta}$, and the field equations are free of the main distributional subtleties we encounter in a generic gauge~\cite{Upton:2021oxf}. But a $1/\rho^2$ divergence appears generically, and practical methods of utilizing the highly regular gauge remain undeveloped.} Here we allow for a mass dipole moment, as in Eq.~\eqref{eq:multipole moments in h}, but restrict to nonspinning bodies. We comment on the generality of our results, outside Lorenz gauge, in the Conclusion.

Introducing a puncture and residual field as at first order, we obtain the effective-source equation
\begin{equation}\label{eq:EFELorenz residual}
    \delta G^{\alpha\beta}[h^{\calR(2)}] = -\Bigl(\delta^2 G^{\alpha\beta}[h^{(1)},h^{(1)}] + \delta G^{\alpha\beta}[h^{\calP(2)}]\Bigr)^\star.
\end{equation}
This is the equation that has been solved in practice in second-order calculations in a Schwarzschild background~\cite{Miller:2023ers,Upton:2025bja}. As at first order, solving this field equation, subject to continuity of the residual field on $\gamma$, yields a total physical field identical to one that would satisfy Eq.~\eqref{eq:EFE2 off gamma} subject to Eq.~\eqref{eq:h2 near gamma}. But unlike at first order, here the original source $\delta^2 G^{\alpha\beta}[h^{(1)},h^{(1)}]$ is strongly divergent, blowing up as $1/\rho^4$ due to the $1/\rho$ blowup of $h^{(1)}_{\alpha\beta}$ and the two derivatives contained in $\delta^2 G^{\alpha\beta}$. Such a singularity is not locally integrable at $\gamma$. It would still be a well-defined distributional source if it took the form of a linear operator ${\cal L}$ acting on an integrable function $f$, since the action of ${\cal L}f$ on a test field $\varphi$ is $\langle \varphi,{\cal L}f\rangle\coloneqq \langle {\cal L}^\dagger\varphi,f\rangle = \int ({\cal L}^\dagger\varphi)f dV$. But $\delta^2 G^{\alpha\beta}$ is a quadratic operator rather than a linear one. A puncture scheme is then seemingly required to obtain a well-defined source on a domain that includes $\gamma$.

Nevertheless, in Ref.~\cite{Upton:2021oxf} two of us showed that one \emph{can} define a distributional source for $h^{(2)}_{\alpha\beta}$, in analogy with the first-order equation~\eqref{eq:EFE1 distributional}. Specifically, taking inspiration from Detweiler~\cite{Detweiler:2011tt}, we showed the total field $h^{(2)}_{\alpha\beta}=h^{\calR(2)}_{\alpha\beta}+h^{\calP(2)}_{\alpha\beta}$ obtained by solving Eq.~\eqref{eq:EFELorenz residual} is identical to the field obtained by solving
\begin{equation}\label{eq:fullEFELorenz}
    \delta G^{\alpha\beta}[h^{(2)}] = 8\pi \tilde T^{\alpha\beta}_{(2)} - \widehat{\delta^2 G}{}^{\alpha\beta}[h^{(1)},h^{(1)}] 
\end{equation}
directly for $h^{(2)}_{\alpha\beta}$. Here, $8\pi\tilde T^{\alpha\beta}_{(2)}\coloneqq \delta G^{\alpha\beta}[h^{(2)}]+\widehat{\delta^2 G}{}^{\alpha\beta}[h^{(1)},h^{(1)}]$ is defined from the local field near $\gamma$, as at first order. It is specifically found to be the second-order term in the expansion of the Detweiler stress-energy tensor,
\begin{equation}
    \tilde T^{\alpha\beta} = m \int_\gamma \tilde{u}^\alpha\tilde{u}^\beta \tilde\delta^4(x,x_p)d\tilde{\tau}, \label{eq:DetT}
\end{equation}
where $\tilde\delta^4(x,x_p)\coloneqq\delta^4(x-x_p)/\sqrt{-\tilde{g}}$, $\tilde u^\alpha\coloneqq dx^\alpha/d\tilde\tau$, and $\tilde\tau$ is proper time in $\tilde g_{\alpha\beta}$. 

The second source term in Eq.~\eqref{eq:fullEFELorenz}, $\widehat{\delta^2 G}{}^{\alpha\beta}[h^{(1)},h^{(1)}]$, is a distributional promotion of ${\delta^2 G}{}^{\alpha\beta}[h^{(1)},h^{(1)}]$, which we termed its ``canonical'' distributional definition in Ref.~\cite{Upton:2021oxf}. To understand why such a distributional promotion must exist, we only need to observe that ${h^{(2)}_{\alpha\beta}= \O(1/\rho^2)}$ is integrable. Therefore, $\delta G^{\alpha\beta}[h^{(2)}]$ is defined as a distribution, and we must be able to construct a distributional source on the right-hand side of Eq.~\eqref{eq:fullEFELorenz}. 

In Ref.~\cite{Upton:2021oxf} we found this distributional source by first noting that, in a neighborhood of $\gamma$, we can decompose the quadratic Einstein tensor into a sum of ``singular-singular'', ``singular-regular'', and ``regular-regular'' terms,
\begin{align}\label{eq:ddG decomp}
    \delta^2G^{\alpha \beta}[h^{(1)},h^{(1)}] &= \delta^2G^{\alpha \beta}[h^{\S(1)},h^{\S(1)}]  + 2Q_\R^{\alpha \beta}[h^{\S(1)}] \nonumber\\
    &\quad + \delta^2G^{\alpha \beta}[h^{\R(1)},h^{\R(1)}],
\end{align}
where we define
\begin{align}\label{eq:Q def}
    Q^{\alpha\beta}_\R[h^{\S(1)}] \coloneqq \delta^2 G^{\alpha\beta}[h^{\R(1)},h^{\S(1)}].
\end{align}
Here we adopt a definition of $\delta^2 G^{\alpha\beta}$ that is bilinear and symmetric in its two arguments when the arguments differ. Since $h^{\R(1)}_{\alpha\beta}$ is smooth, the regular-regular term in Eq.~\eqref{eq:ddG decomp} does not require special treatment. Moreover, $Q^{\alpha\beta}_\R[h^{\S(1)}]$ is a smooth linear operator acting on an integrable function, making it well defined as a distribution too. 
The only term that lacks an immediate distributional definition is $\delta^2G^{\alpha \beta}[h^{\S(1)},h^{\S(1)}]$. But this term is also a linear quantity in disguise, as it must agree 
with $-\delta G^{\alpha\beta}[h^{\rm SS}]$ pointwise, where $h^{\rm SS}_{\alpha\beta}\sim m^2/\rho^2$ is a particular local solution to
\begin{equation}\label{eq:EFEhSS}
    \delta G^{\alpha\beta}[h^{\rm SS}] = - \delta^2G^{\alpha \beta}[h^{\S(1)},h^{\S(1)}] \quad \text{for }x\notin\gamma,
\end{equation}
given explicitly in Ref.~\cite{Pound:2014xva}. The left-hand side of Eq.~\eqref{eq:EFEhSS} is well defined as a distribution since, again, it is a linear operator acting on an integrable function. The core idea is then simply to equate the right-hand side to the left-hand side as a distribution, extending Eq.~\eqref{eq:EFEhSS} to domains including $\gamma$. 

Since the singular and regular fields are only locally constructible, we can only use Eqs.~\eqref{eq:ddG decomp}--\eqref{eq:EFEhSS} in a neighborhood of $\gamma$. We hence adopt a definition that smoothly transitions from the linear operators near $\gamma$ to the ordinary pointwise function away from $\gamma$: 
\begin{align}
    \widehat{\delta^2G}{}^{\alpha\beta} \coloneqq&{} \lim_{s\to0}\Bigl\{\theta^+_s \delta^2G^{\alpha\beta} \nonumber\\
    &+ \theta^-_s \bigl(-\delta G^{\alpha\beta}[h^{\S\S}] + 2Q_\R^{\alpha\beta}[h^{\S(1)}]\bigr)\Bigr\}. \label{eq:d2GHat}
\end{align}
Here $\theta^-_s$ is a smooth window function equal to 1 in an open neighbourhood of the worldline and vanishing outside a small tube $\rho<s$; similarly, $\theta^+_s \coloneqq 1-\theta^-_s$ is equal to 1 outside the tube and vanishes in a neighbourhood of the worldline. Since the regular-regular piece of Eq.~\eqref{eq:ddG decomp} inside the tube contributes nothing when $s\to0$, we need not include it in the second line of Eq.~\eqref{eq:d2GHat}. 

The limit $s\to0$ in Eq.~\eqref{eq:d2GHat} is defined in the distributional sense, meaning it is taken \emph{after} acting on a test field: $\< \varphi_{\alpha\beta},\lim_{s\to0}D^{\alpha\beta}_s\> \coloneqq \lim_{s\to0}\< \varphi_{\alpha\beta}, D^{\alpha\beta}_s\>$ for any net of distributions $D^{\alpha\beta}_s$. In practice, this means one solves Eq.~\eqref{eq:fullEFELorenz} with a finite tube size $s$, defining a solution $h^{s(2)}_{\alpha\beta}$, and then takes the limit $\lim_{s\to0}h^{s(2)}_{\alpha\beta}$. 

Equation~\eqref{eq:d2GHat} is well defined, and by construction it ensures that Eq.~\eqref{eq:fullEFELorenz} yields the correct $h^{(2)}_{\alpha\beta}$ (i.e., the $h^{(2)}_{\alpha\beta}$ consistent with matched expansions). Additionally, we will show that our definitions are consistent with the Bianchi identity, such that the Bianchi identity applied to Eq.~\eqref{eq:fullEFELorenz} yields the conservation equation $\tilde\nabla_\beta \tilde T^{\alpha\beta}=0$, neatly recovering test-body motion in the effective metric. We discuss this further in Sec.~\ref{sec:T2Conservation}.

However, Eq.~\eqref{eq:d2GHat} is not very practical. Solving Eq.~\eqref{eq:fullEFELorenz} using it would require solving the equation for a sequence of values of $s$ and then numerically extrapolating to $s\to0$. Moreover, since Eq.~\eqref{eq:d2GHat} explicitly involves $h^{\S\S}_{\alpha\beta}$, it requires finding at least the most singular part of the puncture even when trying to bypass a puncture scheme. 

Our first goal in this paper will hence be to rewrite the source in the following more useful form:
\begin{equation}\label{eq:ddGhat distribution}
    \widehat{\delta^2G}{}^{\alpha\beta} = \lim_{s\to0}\Bigl[\theta(\rho-s) \delta^2G^{\alpha\beta}-8\pi T^{\alpha\beta}_{\SS}\Bigr]
     + 8\pi T^{\alpha\beta}_Q.
\end{equation}
Here $T^{\alpha\beta}_{\SS}$ and $T^{\alpha\beta}_Q$ are both Dirac $\delta$ functions on $\gamma$, with the schematic forms $\displaystyle\frac{1}{s}\int t^{\alpha\beta}_{\S\S} \delta^4(x,x_p)d\tau$ and $\int h^{\R(1)}_{\mu\nu}t^{\mu\nu\alpha\beta}_Q \delta^4(x,x_p) d\tau$. Such delta sources are easy to deal with using the same methods as at first order. $T^{\alpha\beta}_{\SS}$ acts as a \emph{counterterm}, cancelling the $1/s$ divergence that one would encounter when solving the field equations with the source $\theta(\rho-s) \delta^2G^{\alpha\beta}$. The term $T^{\alpha\beta}_Q$ is a simplification of 
\begin{equation}
T^{\alpha\beta}_Q \coloneqq \frac{1}{4\pi }\lim_{s\to0}\theta^-_s Q_\R^{\alpha\beta}[h^{\rm S(1)}].
\end{equation}
Unlike $T^{\alpha\beta}_{\SS}$, it is finite in the $s\to0$ limit; one would entirely miss it if one were to solve the field equation with a source $\theta(\rho-s) \delta^2G^{\alpha\beta}$ and then simply discard divergent, $1/s$ terms.

The form~\eqref{eq:ddGhat distribution} is an improvement over Eq.~\eqref{eq:d2GHat}, as it allows one to solve field equations with familiar delta-function sources and circumvents the need to compute any part of the puncture field. But it is still not as practical as one might hope. In particular, it still requires solving the field equations for a sequence of $s$ values and then taking the $s\to0$ limit. 

Fortunately, we can establish its equivalence to even more practical ``off the shelf'' regularization schemes. In this paper, we focus on Blanchet and Damour's Hadamard \emph{partie finie} regularization, commonly used in post-Newtonian theory~\cite{Blanchet:1985sp,Blanchet:2000nu,Blanchet:2013haa}. Concretely, we show that Eq.~\eqref{eq:ddGhat distribution} is equivalent to  
\begin{equation} \label{eq:ddG = PfddG}
    \widehat{\delta^2G}{}^{\alpha\beta} = {\rm Pf}\,\delta^2G^{\alpha\beta} + 8\pi T^{\alpha\beta}_Q 
\end{equation}
where the partie finie distribution is defined by 
\begin{equation}
    \< \varphi_{\alpha\beta},{\rm Pf}\,\delta^2G^{\alpha\beta}\> \coloneqq \FP_{B=0}\int (\rho/\rho_0)^B\varphi_{\alpha\beta}\delta^2 G^{\alpha\beta}dV,
\end{equation}
with $\rho_0$ an arbitrary length scale (which must drop out of any final result for a physical field). The ``finite part of a function $f(B)$ at $B=0$'', $\FP_{B=0}f(B)$, denotes the following operation. We first evaluate the function in a region of the complex-$B$ plane where it is analytic in $B$. We then analytically continue to a neighborhood of $B=0$, construct the Laurent series in powers of $B$, and read off the coefficient of $B^0$. That coefficient is the finite part. 

While Hadamard regularization might seem complicated, it is straightforward to implement in practice. We describe its utility in detail in the body of the paper, showing how it enables Green's-function representations of $h^{(2)}_{\alpha\beta}$ and of scalar variables (such as Weyl scalars) constructed from $h^{(2)}_{\alpha\beta}$. Perhaps most importantly, it enables solving directly for $h^{(2)}_{\alpha\beta}$ or associated scalar variables at the level of spherical or spheroidal $\ell m$ modes, which is the traditional method of working in black hole perturbation theory. 

We expect other standard regularization methods, such as dimensional regularization, to be equally applicable and even more broadly accessible in the community outside traditional self-force.


\subsection{Outline}

The layout of the paper is as follows. In Sec.~\ref{sec:local expansions} we review the main tools and definitions needed for our analysis. In Sec.~\ref{sec:EFEsource} we derive Eq.~\eqref{eq:ddGhat distribution} and show its consistency with $\tilde\nabla_\beta \tilde T^{\alpha\beta}=0$, which then implies the equation of motion $\tilde u^\beta\tilde\nabla_\beta \tilde u^\alpha=0$ for the particle's trajectory. Section~\ref{sec:Hadamard} shows the Hadamard regularization~\eqref{eq:ddG = PfddG} is equivalent to Eq.~\eqref{eq:ddGhat distribution}. 

In Sec.~\ref{sec:applications} we explain and demonstrate the practical utility of our results. Specifically, we show how it enables Green's-function representations of solutions, how it facilitates calculations at the level of spherical or spheroidal harmonics in black hole background spacetimes, and how it works naturally with calculations in terms of Teukolsky variables and other master scalars (as opposed to metric variables). Readers interested in these applications but wary of the preceding mathematical gymnastics can safely skip immediately to Sec.~\ref{sec:applications}.

We conclude in Sec.~\ref{sec:discussion} with a discussion of future applications and possible further extensions.

\section{Local expansions in Fermi-Walker coordinates}
\label{sec:local expansions}

Most of our derivations will rely on use of Fermi-Walker coordinates $(\tau,x^i)$ centered on the body's representative worldline $\gamma$, as introduced in the Introduction. We refer to Ref.~\cite{Poisson:2011nh} for a detailed, pedagogical review. Here we briefly summarize the most salient features of the coordinates and the forms that relevant fields take in them. 

In these coordinates, the background metric is given by%
\begin{equation}\label{eq:g in FW}
    g_{\alpha\beta} = \eta_{\alpha\beta} +\O(\e\rho,\rho^2),
\end{equation}
where $\eta={\rm diag}(-1,1,1,1)$. Note that even though this is the external background spacetime, it inherits $\e$ dependence in Fermi-Walker coordinates because it depends on $\gamma$'s acceleration, which is $\e$ dependent. The order-$\rho$ terms in Eq.~\eqref{eq:g in FW} are linear in the acceleration, while the $\rho^2$ terms are proportional to the background spacetime's tidal moments evaluated on $\gamma$. 

To write expressions in compact form, we introduce the four-vector $n^\alpha=(0,n^i)$, where we recall $n^i=x^i/\rho$ is a radial unit vector satisfying $\delta_{ij}n^i n^j =1$. Similarly, we introduce the four-tensor 
\begin{equation}\label{eq:P}
    P^{\alpha\beta} = g^{\alpha\beta} + u^\alpha u^\beta
\end{equation}
on $\gamma$; since $g^{\alpha\beta}|_\gamma = \eta^{\alpha\beta}$ and $u^\alpha=(1,0,0,0)$, this has components $P^{ij}=\delta^{ij}$. We also employ symmetric trace-free index notation, defining $\hat n^{i_1\cdots i_\ell} = n^{\langle i_1}\cdots n^{i_\ell\rangle}$, as well as its natural promotion to four-dimensional indices; for example, $\hat n^{ij} = n^i n^j - \frac{1}{3}\delta^{ij}$, and $\hat n^{\alpha\beta} = n^\alpha n^\beta - \frac{1}{3}P^{\alpha\beta}$. 

We will very frequently have call to evaluate volume integrals in a neighborhood of $\gamma$. The volume element is best written in terms of polar Fermi-Walker coordinates,
\begin{equation}
    dV = \rho^2[1+\O(\e\rho,\rho^2)] d\tau\,d\rho\, d\Omega  \label{eq:FWdV},
\end{equation}
where $d\Omega$ is the surface element of a unit two-sphere. Similarly, we will often evaluate integrals over the surface of a worldtube of constant radius $\rho$, which has surface element
\begin{equation}
    dS^\alpha = -\rho^2[1+\O(\e\rho,\rho^2)]n^\alpha d\tau\,d\Omega.  \label{eq:FWdS}
\end{equation}
Here we note we adopt the convention that this element is directed \emph{inward}. It will be particularly useful to note that the integral $\int f(n^i)d\Omega$ vanishes for any odd-parity function $f$, by which we mean $f(n^i) = - f(-n^i)$. So, in particular,
\begin{equation}
    \int n^{i_1}\cdots n^{i_\ell} d\Omega = 0 \quad \text{for odd }\ell. 
\end{equation}
Moreover, the three-tensors $\hat n^L$ provide a complete basis of orthogonal functions on the two-sphere, meaning any sufficiently regular $f(n^i)$ can be expanded as  $f(n^i) = \sum_{\ell\geq0}f_L \hat n^L$. This is equivalent to an expansion in spherical harmonics, and
\begin{equation}\label{eq:int nhat=0}
    \int \hat n^L d\Omega = 0 \quad \text{for } \ell>0
\end{equation}
in perfect analogy with $\int Y_{\ell m}d\Omega=0$ for all $\ell>0$.

Finally, we will require the first-order singular field, which takes the form
\begin{align}\label{eq:hS1}
    h^{\S(1)}_{\alpha\beta} = \frac{2m(\eta_{\alpha\beta}+2u_\alpha u_\beta)}{\rho} + \O(\e\rho^0,\rho),
\end{align}
and the singular-singular piece of the second-order metric perturbation, which takes the form~\cite{Pound:2014xva}
\begin{align}\label{eq:hSS}
    h^{\S\S}_{\alpha\beta} = \frac{m^2}{3\rho^2}\bigl(8\eta_{\alpha\beta}+2u_\alpha u_\beta -21\hat n_{\alpha\beta} \bigr) +\O(\e\rho^{-1},\rho^0).
\end{align}
In these expressions, the $\O(\e)$ terms arise from dependence on the acceleration, as for the background metric. In practice, following Ref.~\cite{Upton:2021oxf}, we can demote these acceleration terms to higher-order pieces of the metric. For example, we can move the $\O(\e\rho)$ term in $g_{\alpha\beta}$ into $h^{\R(1)}_{\alpha\beta}$; the $\O(\e\rho^0)$ term in $h^{\S(1)}_{\alpha\beta}$, into $h^{\S\R}_{\alpha\beta}$, where it is not sufficiently singular to enter our derivations; and the $\O(\e/\rho)$ term in $h^{\S\S}_{\alpha\beta}$, into $h^{(3)}_{\alpha\beta}$.

We will only require one feature of higher-order-in-$\rho$ terms in the above expressions: each successive order has definite parity, and the parity reverses order by order. So, for example, the order-$\rho^k$ terms in $h^{\S(1)}_{\alpha\beta}$ have odd parity for even $k$ and even parity for odd $k$, while the order-$\rho^k$ terms in $h^{\S\S}_{\alpha\beta}$ have even parity for even $k$ and odd parity for odd $k$. The singular-regular piece of the perturbation, $h^{\S\R}_{\alpha\beta}$, has the same parity as $h^{\S(1)}_{\alpha\beta}$ at each order in $\rho$ (and therefore the opposite parity from $h^{\S\S}_{\alpha\beta}$).

Finally, we will need the leading terms in the second-order quadratic source. The singular-singular term in Eq.~\eqref{eq:ddG decomp} is
\begin{multline}
    \delta^2G^{\alpha \beta}[h^{\S(1)},h^{\S(1)}] = \frac{m^2}{3\rho^4}(2u^\alpha u^\beta + 42\hat{n}^{\alpha \beta}-7\eta^{\alpha \beta})  \\
    +\O(\e\rho^{-3}),\label{eq:d2GSSexpansion}
\end{multline}
which is most easily derived using $\delta^2G^{\alpha \beta}[h^{\S(1)},h^{\S(1)}] = -\delta G^{\alpha\beta}[h^{\S\S}]+\O(\rho^{-2})$ and substituting Eq.~\eqref{eq:hSS}.
The singular-regular term in Eq.~\eqref{eq:ddG decomp} is
\begin{multline}\label{eq:QR near gamma}
2Q_\R^{\alpha\beta}[h^{\S(1)}] = \frac{k^{\alpha\beta}{}_{\mu\nu}\hat{n}^{\mu\nu}}{\rho^3} + \frac{j^{\alpha\beta}{}_{\mu}n^{\mu}+j^{\alpha\beta}{}_{\mu\nu\sigma}\hat n^{\mu\nu\sigma}}{\rho^2} \\
+ \mathcal{O}(\rho^{-1}),
\end{multline}
where 
\begin{align}\label{eq:QR rho3 coeff}
    k^{\alpha\beta}{}_{\mu\nu} ={}& 3m\Bigl[h^{\R(1)}_{\mu\nu}(P^{\alpha\beta}+u^\alpha u^\beta) \nonumber \\
        & - 2h^{\R(1)}_{\rho(\mu}\delta_{\nu)}^{(\alpha}(P^{\beta)\rho}+u^{\beta)}u^{\rho}) \nonumber \\
        & + \delta^{(\alpha}_{(\mu}\delta^{\beta)}_{\nu)}h^{\R(1)}_{\rho\sigma}(P^{\rho\sigma} + u^\rho u^\sigma)\Bigr]\Bigr|_\gamma.
\end{align}
Equations~\eqref{eq:d2GSSexpansion}--\eqref{eq:QR rho3 coeff} correspond to trace-reversals of Eqs.~(D3)--(D6) from Ref.~\cite{Pound:2009sm}. We omit the lengthy expressions for  $j^{\alpha\beta}{}_{\mu}$ and $j^{\alpha\beta}{}_{\mu\nu\sigma}$, both of which are made up of linear combinations of first derivatives of $h^{\R(1)}_{\alpha\beta}$; these can be extracted from Eqs.~(D7)--(D9) of Ref.~\cite{Pound:2009sm}. 

Like the metric perturbations, higher-order terms in $\delta^2G^{\alpha\beta}[h^{\S(1)},h^{\S(1)}]$ and $Q_\R^{\alpha\beta}[h^{\S(1)}]$ alternate between even and odd parity at successive orders in $\rho$.

\section{Distributional source for the second-order Einstein equation}\label{sec:EFEsource}

In this section, we derive Eq.~\eqref{eq:ddGhat distribution}, the counterterm version of $\widehat{\delta^2G}{}^{\alpha\beta}$. We then show that this definition implies conservation of the effective stress-energy skeleton, Eq.~\eqref{eq:TDet conservation}, which in turn implies the test-particle equation of motion $\tilde u^\beta\tilde\nabla_\beta\tilde u^\alpha=0$ (up to second-order terms and recalling that we neglect spin and higher multipole).

In this and later sections, we work with compactly supported test fields unless stated otherwise.

\subsection{Counterterm derivation}

Equation~\eqref{eq:fullEFELorenz} is well defined with our distributional definition~\eqref{eq:d2GHat}. But it is impractical for the reasons described in the Introduction: it requires directly using part of the puncture field even when not employing a puncture scheme, and it requires solving the field equations for a sequence of $s$ values before taking the limit $s\to 0$. Recasting~\eqref{eq:d2GHat} in the counterterm form~\eqref{eq:ddGhat distribution} entirely bypasses the first of these issues and ameliorates the second.

First we note that Eq.~\eqref{eq:d2GHat} implies
\begin{align}
    \hspace{-3pt}\langle\varphi_{\alpha \beta}, \widehat{\delta^2G}{}^{\alpha \beta}\rangle \coloneqq \lim_{s \to 0}& \biggl\{\int\theta^+_s \delta^2G^{\alpha\beta} \varphi_{\alpha \beta} dV \nonumber\\
    &-\int_{\rho<s} h^{\rm SS}_{\alpha \beta} \delta G^{\alpha\beta}[\theta^-_s\varphi] dV\nonumber
    \\
    &+2\int_{\rho<s}  h^{\rm S(1)}_{\alpha \beta} Q_{\R}^{\dagger\alpha\beta}[ \theta^-_s\varphi] dV\biggr\} ,\label{eq:<phi,ddGhat>}
\end{align}
where we have used the facts that \(\deltaG_{\alpha\beta}\) is self-adjoint and that $\theta^-_s = 0$ for $\rho>s$. Next, we split up the first integral:
\begin{align}
    &\int\theta^+_s \delta^2G^{\alpha\beta} \varphi_{\alpha \beta} dV \nonumber\\
    &=\int_{\rho>s} \delta^2G^{\alpha\beta} \varphi_{\alpha \beta} dV 
    + \int_{\rho<s}\theta^+_s \delta^2G^{\alpha\beta} \varphi_{\alpha \beta} dV \nonumber
    \\
    &=\int_{\rho>s} \delta^2G^{\alpha\beta} \varphi_{\alpha \beta} dV +\int_{\rho<s} \theta^+_s \bigl(-\delta G^{\alpha\beta}[h^{\rm SS}]\bigr) \varphi_{\alpha \beta} dV\nonumber
    \\
    &+\int_{\rho<s} \theta^+_s \bigl( 2Q_\R^{\alpha\beta}[h^{\rm S(1)}] + \delta^2G^{\alpha\beta}[h^{\R (1)},h^{\R (1)}]\bigr) \varphi_{\alpha \beta} dV,\label{eq:int theta+ddGphi}
\end{align}
where in the first equality we used the fact that $\theta^+_s =1$ for $\rho>s$. For the second equality we exploited the decomposition~\eqref{eq:ddG decomp} and the equality ${\delta^2 G^{\alpha\beta}[h^{\S(1)},h^{\S(1)}]=-\delta G^{\alpha\beta}[h^{\S\S}]}$. 
Combining the above results yields 
\begin{align}
    \langle\varphi_{\alpha \beta}, \widehat{\delta^2G}{}^{\alpha \beta}\rangle = \lim_{s \to 0}& \biggl\{\int_{\rho>s} \delta^2G^{\alpha\beta} \varphi_{\alpha \beta} dV \nonumber
    \\
    &+\int_{\rho<s} \theta^+_s \bigl(-\delta G^{\alpha\beta}[h^{\rm SS}]\bigr) \varphi_{\alpha \beta} dV\nonumber
    \\
    &+\int_{\rho<s} \theta^+_s \bigl( 2Q_\R^{\alpha\beta}[h^{\rm S(1)}]\bigr) \varphi_{\alpha \beta} dV\nonumber
    \\
    &-\int_{\rho<s}  h^{\rm SS}_{\alpha \beta} \deltaG^{\alpha\beta}[ \varphi^s] dV\nonumber
    \\
    &+\int_{\rho<s}  2h^{\rm S(1)}_{\alpha \beta} Q_{\R}^{\dagger\alpha\beta}[\varphi^s] dV\biggr\}, 
    \label{eq31}
\end{align}
with $\varphi^s_{\alpha\beta} \coloneqq \varphi_{\alpha\beta} \theta^-_s$. Here we dropped the regular-regular term from Eq.~\eqref{eq:int theta+ddGphi} because it manifestly contributes nothing in the $s\to0$ limit. One might guess that the final term in Eq.~\eqref{eq31} also vanishes in that limit because it behaves as  $\int_{\rho<s} Q_{\R}^{\dagger\alpha\beta}[\varphi^s]\rho d\tau d\rho d\Omega$. However, our derivation below shows it is nonzero when $s\to0$; this is because $Q_{\R}^{\dagger\alpha\beta}[\varphi^s]$ contains derivatives of $\theta^-_s$, which diverge as $s\to0$ because the window function becomes infinitely steep in the limit.

To calculate the contribution of the \(h^{\S\S}_{\alpha \beta}\) terms in Eq.~\eqref{eq31}, we follow the same procedure as in Refs.~\cite{Upton:2021oxf,Upton:2025bja} when determining the form of the stress-energy tensor. We start by integrating the second-to-last term in \eqref{eq31}:
\begin{align}
	\MoveEqLeft[4] -\int_{\rho<s}h^{\S\S}_{\alpha\beta}\deltaG^{\alpha\beta}[\varphi^s]dV\nonumber \\
        ={}& -\lim_{R\to 0^+}\int_{R<\rho<s}h^{\S\S}_{\alpha\beta}\deltaG^{\alpha\beta}[\varphi^s]dV\nonumber\\
        ={}& \lim_{R\to 0^+}\Bigl(-\int_{R<\rho<s}\deltaG^{\alpha\beta}[h^{\S\S}]\varphi_{\alpha\beta}^sdV \nonumber \\
            & + \int_{\rho=R}K^{\deltaG}_{\alpha}[\varphi^s,h^{\S\S}]dS^{\alpha}\Bigr), \label{eq:dGhSSDistInt}
\end{align}
where $K_{\alpha}^{\delta G}[\varphi,h]$ is defined by 
\begin{equation}
\nabla^\alpha K_{\alpha}^{\delta G}[\varphi,h]\coloneqq \deltaG^{\alpha\beta}[h]\varphi_{\alpha\beta}-\deltaG^{\alpha\beta}[\varphi]h_{\alpha\beta}.    
\end{equation}
In the first line of \eqref{eq:dGhSSDistInt} we used the fact that $\int f dV = \lim_{R\to0^+}\int_{\rho>R}fdV$ for any integrable $f$. In the second line we integrated by parts, noting that the boundary term at $\rho=s$ vanishes because $\varphi^s_{\alpha\beta}$ vanishes there. Unlike in similar calculations in our previous work~\cite{Upton:2021oxf,Upton:2025bja}, the volume integral in Eq.~\eqref{eq:dGhSSDistInt} does not vanish as the integrand is nonzero off the worldline.
This means we must explicitly calculate its contribution to the final result. 

First calculating the boundary term in Eq.~\eqref{eq:dGhSSDistInt}, we appeal to the explicit expression 
\begin{align}
    K_{\alpha}^{\delta G}[\varphi,h] &= -\frac{1}{2} \varphi^{\beta \mu} \nabla_\alpha h_{\beta \mu} +\frac{1}{2} h^{\beta \mu} \nabla_\alpha \varphi_{\beta \mu} \nonumber \\
        & \quad + \varphi^{\beta}{}_{\beta} \nabla_{[\alpha} h^\mu{}_{\mu]} -h^\beta{}_{\beta} \nabla_{[\alpha} \varphi^\mu{}_{\mu]} \nonumber \\
        & \quad - \frac{1}{2} \varphi_\alpha{}^\beta \nabla_{\beta} h^{\mu}{}_{\mu} +  \frac{1}{2} h_\alpha{}^\beta \nabla_{\beta} \varphi^{\mu}{}_{\mu} \nonumber
    \\
        & \quad -h^{\beta \mu} \nabla_{\mu} \varphi_{\alpha \beta} + \varphi^{\beta \mu} \nabla_\mu h_{\alpha\beta}\,, \label{eq:KdeltaGdef}
\end{align}
which is taken from Eq.~(214) of Ref.~\cite{Upton:2021oxf} (with the correction of an overall minus sign).
We substitute \(h^{\S\S}_{\alpha\beta}\) from Eq.~\eqref{eq:hSS} to find that
\begin{align}
    K^{\deltaG}_{\alpha}[\varphi^s,h^{\SS}] ={}& \frac{m^2\varphi_{\beta\mu}^s}{\rho^3}(10\delta_\alpha^{(\mu} n^{\beta)} + 14n_\alpha{}^{\beta\mu}-10n^{(\beta} P_\alpha{}^{\mu)} \nonumber \\
        & - 7n_\alpha P^{\beta\mu} + 4u_{\alpha}n^{(\beta}u^{\mu)} + 3n_{\alpha}u^{\beta}u^{\mu}) \nonumber \\
        & + \O(\rho^{-2}). \label{eq:kdGhSS}
\end{align}
While terms of order $\rho^{-2}$ do appear in $K^{\deltaG}_{\alpha}[\varphi,h^{\SS}]$, we can safely ignore them as, due to the parity of the singular field in the Lorenz gauge, each term will vanish when performing the angular integration.
We evaluate the surface integral by contracting Eq.~\eqref{eq:kdGhSS} with the Fermi-Walker surface element~\eqref{eq:FWdS} and integrating over the angles to find that
\begin{multline}
	\int_{\rho=R}K^{\deltaG}_{\alpha}[\varphi^s,h^{\S\S}]dS^{\alpha} \\
        = \frac{4\pi m^2}{3} \int_\gamma \frac{\varphi_{\alpha\beta}}{R}(7P^{\alpha\beta}-9u^{\alpha}u^{\beta})d \tau  + \O(R), \label{eq:kdGhSSAngInt} 
\end{multline}
where we have replaced $\varphi^s_{\alpha\beta}$ with $\varphi_{\alpha\beta}$ as the integral is performed along the worldline and $\varphi^s_{\alpha\beta}|_{\gamma}=\varphi_{\alpha\beta}|_{\gamma}$.

Returning to the volume integral in Eq.~\eqref{eq:dGhSSDistInt}, we combine it with the volume integral in \eqref{eq31}, use $\theta^+_s +\theta^-_s =1$, and expand in Fermi-Walker coordinates using Eqs.~\eqref{eq:FWdV} and~\eqref{eq:hSS}. The result is
\begin{align}
    \MoveEqLeft[2]\int_{R<\rho<s}\deltaG^{\alpha\beta}[h^{\S\S}]\varphi_{\alpha\beta}dV \nonumber \\
        ={}& \int_{R<\rho<s}\biggl[\frac{m^2\varphi_{\alpha\beta}}{3\rho^4}(7P^{\alpha\beta}-9u^{\alpha}u^{\beta}-42\nhat^{\alpha\beta}) \nonumber\\
        &\qquad\qquad +\O(\rho^{-3})\biggr]dV \nonumber \\
        ={}& \int_{R<\rho<s}\left[\frac{4m^2\pi\varphi_{\alpha\beta}}{3\rho^2}(7P^{\alpha\beta}-9u^{\alpha}u^{\beta}) + \O(\rho^{0})\right]d \tau d\rho \nonumber \\
        ={}& \frac{4m^2\pi}{3Rs}(s-R)\int_\gamma\varphi_{\alpha\beta}(7P^{\alpha\beta}-9u^\alpha u^\beta)d\tau + \O(R,s), \label{eq:dGhSSIntegrand}
\end{align}
where the order-\(\rho^{-1}\) term in the second equality vanishes under angular integration going from the first to second equality due to its odd parity.

Combining Eqs.~\eqref{eq:dGhSSDistInt}, \eqref{eq:kdGhSSAngInt}, and~\eqref{eq:dGhSSIntegrand}, and taking the limit as $R\to 0^+$, we find that the $h^{\S\S}_{\alpha\beta}$ terms in Eq.~\eqref{eq31} reduce to
\begin{align}
    &\int_{\rho<s} \theta^+_s \bigl(-\delta G^{\alpha\beta}[h^{\rm SS}]\bigr) \varphi_{\alpha \beta} dV -\int_{\rho<s}  h^{\rm SS}_{\alpha \beta} \deltaG^{\alpha\beta}[ \varphi^s] dV\nonumber
    \\
    &=\frac{4\pi m^2}{3s}\int_\gamma (7g^{\alpha\beta}-2u^{\alpha}u^{\beta})\varphi_{\alpha\beta}d\tau, \label{eq:CounterTerm}
\end{align}
where we have written the projection operator $P^{\alpha\beta}$ in terms of the metric and four-velocity. We can think of Eq.~\eqref{eq:CounterTerm} as integration of $\varphi_{\alpha\beta}$ against ($-8\pi$ times) an effective stress-energy tensor, $-8\pi\langle\varphi_{\alpha\beta}, T^{\alpha\beta}_{\S\S}\rangle$, with 
\begin{equation}
	T_{\SS}^{\alpha\beta} = -\frac{m^2}{6s}\int_\gamma (7g^{\alpha\beta}-2u^{\alpha}u^{\beta})\delta^{4}(x,x_p)d\tau. \label{eq:Tcount}
\end{equation}
Readers should observe that the integrand, up to a constant factor, is simply the angle average of the coefficient of $1/\rho^4$ in Eq.~\eqref{eq:d2GSSexpansion}. In other words, the counterterm essentially removes the angle average of the most singular term in the source.

Having completed the calculation of \(h^{\S\S}_{\alpha\beta}\) contributions, we turn our attention to the singular-regular pieces in Eq.~\eqref{eq31}. First consider the third term in Eq.~\eqref{eq31}. This term is $\O(s^2)$ due to the angular structure of the integrand. Referring to Eq.~\eqref{eq:QR near gamma}, we see the angular integral of the $1/\rho^3$ term vanishes by virtue of Eq.~\eqref{eq:int nhat=0}, and the angular integral of $1/\rho^2$ terms vanishes because they have odd parity.

Next consider the last term in \eqref{eq31}. Following the same steps as in Eq.~\eqref{eq:dGhSSDistInt}, we obtain
\begin{align}
	\MoveEqLeft[4] \int_{\rho<s} Q_{\R}^{\dagger\alpha\beta}[\varphi^s]h^{\S(1)}_{\alpha\beta}dV \nonumber \\
        ={}& \lim_{R\to 0^+}\biggl(\int_{R<\rho<s}\varphi_{\alpha\beta}^sQ_{\R}^{\alpha\beta}[h^{\S(1)}]dV \nonumber \\
	           & - \int_{\rho=R}K^{Q}_{\alpha}[\varphi^s,h^{\S(1)}]dS^{\alpha}\biggr), \label{eq:Qdist}
\end{align}
where $K^Q_\alpha[\varphi,h^{\S(1)}]$ is defined by 
\begin{equation}
\nabla^\alpha K_{\alpha}^{Q}[\varphi,h]\coloneqq Q_\R^{\alpha\beta}[h]\varphi_{\alpha\beta}-Q_\R^{\dagger\alpha\beta}[\varphi]h_{\alpha\beta}    
\end{equation}
and given explicitly by Eq.~(222) in Ref.~\cite{Upton:2021oxf}. The volume integral in Eq.~\eqref{eq:Qdist} vanishes in the $R,s\to0$ limit by virtue of Eq.~\eqref{eq:QR near gamma}. So we are only left with the boundary term, which is given by Eq.~(227) in Ref.~\cite{Upton:2021oxf}:\footnote{Equation~(227) of Ref.~\cite{Upton:2021oxf} is missing a factor of $1/2$ on the right-hand side of the expression, which we have corrected here.}
\begin{align}
    \MoveEqLeft[2] \lim_{R\to 0^+} \int_{\rho=R}K^{Q}_{\alpha}[\varphi^s,h^{\S(1)}]dS^{\alpha}\nonumber
    \\
    &= \frac{2\pi m}{3}\int\Big(h^{\mathrm{R}(1)}_{ab}\varphi^{ab}
   + 2h^{\mathrm{R}(1)}_{ta}\varphi^{a}{}_{t}+ 2h^{\mathrm{R}(1)}_{ta}\varphi_{t}{}^{a}
    \nonumber\\
&\qquad\qquad\qquad - 2h^{\mathrm{R}(1)}_{tt}\varphi^{a}{}_{a}- \delta^{ij}h^{\mathrm{R}(1)}_{ij}\varphi^{b}{}_{b} \nonumber
   \\
   &\qquad \qquad\qquad+ 4\delta^{ij}h^{\mathrm{R}(1)}_{ij}\varphi_{tt}
   - 6h^{\mathrm{R}(1)}_{tt}\varphi_{tt}\Big)d\tau ,
\end{align}
where we once again used the fact that $\varphi^s|_\gamma = \varphi|_\gamma$. 
We can write this in terms of a $\delta$ function as
\begin{align}\label{eq:<phi,TQ>}
    \int_{\rho<s} 2h^{\S(1)}_{\alpha\beta}Q_{\R}^{\dagger\alpha\beta}[\varphi^s] dV  = 8 \pi \langle \varphi_{\alpha\beta}, T_Q^{\alpha\beta}\rangle
\end{align}
with
\begin{align}
	T_{Q}^{\alpha\beta} &= -\frac{m}{6}\int h^{R(1)}_{\mu\nu}X^{\alpha\beta\mu\nu}\delta^4(x,x_p) d{\tau}, \label{eq:TQ}
\end{align}
where
\begin{align}\label{eq:X}
    X^{\alpha\beta\mu\nu} \coloneqq{}& P^{\mu(\alpha}P^{\beta)\nu} + 4P^{\mu(\alpha}u^{\beta)}u^{\nu} - 2u^{\mu}u^{\nu}P^{\alpha\beta} \nonumber \\
    & - P^{\mu\nu}P^{\alpha\beta} + 4P^{\mu\nu}u^{\alpha}u^{\beta} - 6u^\mu u^{\nu}u^{\alpha}u^{\beta}.
\end{align}
Here we used Eq.~\eqref{eq:P} and the surrounding identities to express $T_Q^{\alpha \beta}$ in terms of $P^{\alpha \beta}$ and $u^\alpha$.

We now gather our results. Substituting Eqs.~\eqref{eq:CounterTerm} and~\eqref{eq:<phi,TQ>} into Eq.~\eqref{eq31}, along with the vanishing of the third term, we obtain our desired counterterm form of $\widehat{\delta^2G}{}^{\alpha\beta}$, Eq.~\eqref{eq:ddGhat distribution}. The counterterm $T^{\alpha\beta}_{\S\S}$ is given by Eq.~\eqref{eq:Tcount}; the finite term $T^{\alpha\beta}_{Q}$, by Eq.~\eqref{eq:TQ}.

\subsection{Equations of motion and stress-energy conservation}\label{sec:T2Conservation}

The Bianchi identity is a central feature of GR and modified theories of gravity~\cite{Gralla:2010cd}: it ensures the field equations enforce conservation of a body's stress-energy tensor, and the vacuum Bianchi identity is what ensures that the field equations \emph{outside} a body encode the body's equations of motion~\cite{Weyl:1921,Eddington:1924,Einstein:1938yz,Einstein:1949,Infeld:1949}. In this section, we show that our distributional second-order Einstein tensor does satisfy an appropriate distributional Bianchi identity, which implies the conservation of the Detweiler stress-energy tensor in the effective spacetime. 

To understand what the correct Bianchi identity is, first consider the simpler case of a smooth metric ${\sf g}_{\alpha\beta}=g_{\alpha\beta}+ h_{\alpha\beta}$. If $g_{\alpha\beta}$ is a vacuum metric, then the Einstein equations expanded up to second order in $h_{\alpha\beta}$ are given by 
\begin{equation}
    \deltaG^{\alpha\beta}[h] = 8\pi T^{\alpha\beta} - \delta^{2}G^{\alpha\beta}[h,h] + \O(|h|^3). \label{eq:EFEs summed}
\end{equation}
The exact Bianchi identity is ${}^{\sf g}\nabla_\beta G^{\alpha\beta}[\sf g]=0$, where ${}^{\sf g}\nabla_\beta$ is compatible with ${\sf g}_{\alpha\beta}$. Expanding this exact identity implies linear and quadratic ones:
\begin{align}
    \nabla_\beta \deltaG^{\alpha\beta}[h] &= 0,\label{eq:Bianchi 1st order}\\
    \nabla_\beta \delta^2G^{\alpha\beta}[h,h] &= -\delta\Gamma^{\alpha}_{\beta\gamma}[h]\deltaG^{\gamma\beta}[h] \nonumber\\
    &\quad - \delta\Gamma^{\beta}_{\beta\gamma}[h]\deltaG^{\alpha\gamma}[h],\label{eq:Bianchi 2nd order smooth}
\end{align}
where $\delta\Gamma^{\alpha}_{\beta\gamma}$ is the linearized difference between the Christoffel symbols in \({\sf g}_{\mu\nu}\) and \(g_{\mu\nu}\), given by
\begin{equation}\label{eq:dGamma}
    \delta\Gamma^{\alpha}_{\beta\gamma}[h] = \frac{1}{2}g^{\alpha\mu}\bigl(2\nabla_{(\beta}h_{\gamma)\mu}-\nabla_\mu h_{\beta\gamma}\bigr).
\end{equation}
Equations~\eqref{eq:Bianchi 1st order} and \eqref{eq:Bianchi 2nd order smooth} hold for any smooth $h_{\alpha\beta}$. Taking the divergence of Eq.~\eqref{eq:EFEs summed} with respect to the background covariant derivative, substituting Eqs.~\eqref{eq:Bianchi 1st order} and \eqref{eq:Bianchi 2nd order smooth}, and then substituting $\deltaG^{\alpha\beta}=8\pi T^{\alpha\beta} + \O(|h|^2)$, we obtain the second-order conservation equation:
\begin{align}
    0 &= \nabla_\beta T^{\alpha\beta} + \delta\Gamma^{\alpha}_{\beta\gamma}[h]T^{\gamma\beta} + \delta\Gamma^{\beta}_{\beta\gamma}[h]T^{\alpha\gamma} +\O(|T||h|^2) \nonumber\\
    &= {}^{\sf g}\nabla_\beta T^{\alpha\beta}.
\end{align}

We now wish to show that an analogous procedure holds for our distributional field equations. Our analogue of Eq.~\eqref{eq:EFEs summed} is\footnote{In this section we deliberately work with the \emph{total} Einstein equation rather than one equation for $h^{(1)}_{\alpha\beta}$ and one for $h^{(2)}_{\alpha\beta}$. The reason is that the division into the sequence of equations~\eqref{eq:EFE1 distributional} and~\eqref{eq:fullEFELorenz} is only possible in one of two ways: (i) the particle's trajectory is expanded in powers of $\e$, as in $x^\alpha_p = x^\alpha_{(0)}(\tau)+\e x^\alpha_{(1)}(\tau)+\O(\e^2)$, where $x^\alpha_{(0)}$ is a background geodesic and $x^\alpha_{(1)}$ enters as a mass dipole moment in $\tilde T^{\alpha\beta}_{(2)}$ and $h^{(2)}_{\alpha\beta}$ (the ``Gralla-Wald'' approach~\cite{Gralla:2008fg}); or (ii) the operator $\deltaG^{\alpha\beta}$ on the left-hand side of the equations is actually the gauge-fixed Lorenz-gauge version of $\deltaG^{\alpha\beta}$, call it ${\cal E}^{\alpha\beta}$, in which case $x^\alpha_p$ can be the particle's self-accelerated center-of-mass trajectory (the ``self-consistent'' approach~\cite{Pound:2009sm}). In the second approach, the operator does not satisfy the linearized Bianchi identity: $\nabla_\beta{\cal E}^{\alpha\beta}$ is not identically zero. By working with the total equation~\eqref{eq:EFEsSource} we can remain agnostic as to whether the expansion is in Gralla-Wald form or self-consistent form. See Refs.~\cite{Pound:2009sm,Pound:2015fma,Pound:2015tma} for detailed discussion. Section~\ref{sec:applications} briefly reviews the form of the equations used in practice for binary systems.}
\begin{equation}
    \deltaG^{\alpha\beta}[h] = 8\pi \tilde T^{\alpha\beta} - \widehat{\delta^{2}G}{}^{\alpha\beta}[h,h] + \O(\e^3) \,, \label{eq:EFEsSource}
\end{equation}
with $h_{\alpha\beta}=\sum_{n\geq1}\e^n h^{(n)}_{\alpha\beta}$. Equation~\eqref{eq:Bianchi 1st order} actually holds for any distributional $h_{\alpha\beta}$, meaning that taking the divergence of \eqref{eq:EFEsSource} leads to 
\begin{equation}
    8\pi \nabla_{\beta}\Teff^{\alpha\beta} - \nabla_{\beta}\widehat{\delta^{2}G}{}^{\alpha\beta}[h,h] = \O(\e^3). \label{eq:ConservationCondition}
\end{equation}
We would like this to imply conservation of the effective stress-energy tensor in the effective metric, ${\tilde\nabla_\beta\tilde T^{\alpha\beta}=0}$, since that equation implies test-mass motion in the effective metric, $\tilde u^\beta\tilde\nabla_\beta \tilde u^\alpha=0$, which is known to be true (from the field equations outside the body~\cite{Pound:2017psq}) up to unknown $\O(\e^3)$ terms. That conservation equation reads
\begin{align}
    0&=\nabla_{\beta}\Teff^{\alpha\beta} +\delta\tilde\Gamma^{\alpha}_{\beta\gamma}\Teff^{\beta\gamma} + \delta \tilde{\Gamma}^{\beta}_{\beta\gamma}\Teff^{\alpha\gamma}+\O(\e^3)\nonumber\\
    &= \tilde\nabla_\beta \tilde T^{\alpha\beta},\label{eq:DivBGTEff}
\end{align}
where \(\delta\tilde\Gamma^{\alpha}_{\beta\gamma}\coloneqq\delta\Gamma^{\alpha}_{\beta\gamma}[h^{\R}]\) and  $h^\R_{\alpha\beta}=\sum_{n\geq1}\e^n h^{\R(n)}_{\alpha\beta}$. For Eq.~\eqref{eq:ConservationCondition} to imply the correct conservation equations, we require the Bianchi identity for smooth fields, Eq.~\eqref{eq:Bianchi 2nd order smooth}, to be replaced by the following nontrivial distributional Bianchi identity:
\begin{align}\label{eq:Bianchi 2nd order hat}
    \nabla_\beta \widehat{\delta^2G}{}^{\alpha\beta} &= -\delta\tilde\Gamma^{\alpha}_{\beta\gamma}\deltaG^{\beta\gamma}[h^{(1)}]  - \delta\tilde\Gamma^{\beta}_{\beta\gamma}\deltaG^{\alpha\gamma}[h^{(1)}].
\end{align}

Before proving Eq.~\eqref{eq:Bianchi 2nd order hat}, we motivate why it should be true. We have effectively defined the most singular part of $\delta^2G^{\alpha\beta}$ to be $-\deltaG^{\alpha\beta}[h^{\S\S}]$, which satisfies the Bianchi identity $\nabla_\beta\deltaG^{\alpha\beta}[h^{\S\S}]=0$.\footnote{There is a slight subtlety here if one uses the self-consistent rather than Gralla-Wald expansion (see footnote 5). If we use the Lorenz-gauge-fixed operator ${\cal E}^{\alpha\beta}[h^{\S\S}]$ in place of $\deltaG^{\alpha\beta}[h^{\S\S}]$, then it is not exactly conserved. Instead, $\nabla_\beta {\cal E}^{\alpha\beta}[h^{\S\S}]$ contains order-$\e$ acceleration terms. Since these are higher order in $\e$, they do not affect our arguments here. Similarly, ${\cal E}^{\alpha\beta}[h^{\S\S}]$ and $\deltaG^{\alpha\beta}[h^{\S\S}]$ differ from one another by order-$\e$ acceleration terms that are immaterial to our derivations throughout the paper.} Hence, in generalizing Eq.~\eqref{eq:Bianchi 2nd order smooth} we will not have any singular-singular terms on the right-hand side because they identically vanish on the left. Since $h^{\R(1)}_{\alpha\beta}$ is a vacuum solution, $\deltaG^{\alpha\beta}[h^{\R(1)}]$ terms in Eq.~\eqref{eq:Bianchi 2nd order smooth} will also vanish, meaning we will not obtain terms of the form $\delta\Gamma^{\alpha}_{\beta\gamma}[h^{\S(1)}]\deltaG^{\gamma\beta}[h^{\R(1)}]$, for example. The only remaining terms we can expect from Eq.~\eqref{eq:Bianchi 2nd order smooth} are of the form in Eq.~\eqref{eq:Bianchi 2nd order hat}, noting $\deltaG^{\alpha\beta}[h^{(1)}] =\deltaG^{\alpha\beta}[h^{\S(1)}]$.

To set the stage for the rigorous proof, we simplify the right-hand side of Eq.~\eqref{eq:Bianchi 2nd order hat}. Using Eq.~\eqref{eq:dGamma} together with $\deltaG^{\alpha\beta}[h^{(1)}] = 8\pi \tilde T^{\alpha\beta}_{(1)} + \O(\e)$, we find that (up to an overall factor of $-8\pi$), the right-hand side of Eq.~\eqref{eq:Bianchi 2nd order hat} is
\begin{equation} 
    2\delta\tilde\Gamma^{(\alpha}_{\beta\gamma}\tilde T_{(1)}^{\beta)\gamma} = \frac{m}{2} \int_{\gamma}U^{\alpha\beta\gamma\mu}h^{\R(1)}_{\beta\gamma;\mu}\delta^4(x,x_p)d\tau + \O(\e^3), \label{eq:divTtilde}
\end{equation}
where
\begin{equation}
    U^{\alpha\beta\gamma\mu} = 2g^{\alpha(\beta}u^{\gamma)}u^\mu + P^{\beta\gamma}u^{\alpha}u^{\mu} - P^{\alpha\mu}u^\beta u^\gamma. \label{eq:Udef}
\end{equation}
Our goal is therefore to show that $ \nabla_\beta \widehat{\delta^2G}{}^{\alpha\beta}$ reduces to Eq.~\eqref{eq:divTtilde} (up to a factor of $-8\pi$).

The desired equality is a distributional one, meaning we should show it holds at the level of actions on a test field. Our result~\eqref{eq:ddGhat distribution} implies 
\begin{align}
    \hspace{-7pt}\bigl\langle \varphi_\alpha, \nabla_\beta \widehat{\delta^2G}{}^{\alpha\beta} \bigr\rangle &\coloneqq \bigl\langle -\nabla_\beta\varphi_\alpha, \widehat{\delta^2G}{}^{\alpha\beta} \bigr\rangle \\
    &\hphantom{:}= -\lim_{s\to0}\biggl\{\int_{\rho>s} \!\!\delta^{2}G^{\alpha\beta}[h^{(1)},h^{(1)}]\nabla_{\beta}\varphi_{\alpha}dV\nonumber\\
    &\hphantom{:}\ \qquad -8\pi\bigl\langle \nabla_\beta\varphi_\alpha, T^{\alpha\beta}_{\S\S}-T^{\alpha\beta}_{Q}\bigr\rangle\biggr\}.\label{eq:Bianchi on test function}
\end{align}
Integrating the first term by parts and using the vacuum Bianchi identity $\nabla_\beta\delta^{2}G^{\alpha\beta}[h^{(1)},h^{(1)}]=0$ for points away from $\gamma$, we find
\begin{multline}\label{eq:Bianchi IBP}
    \int_{\rho>s} \!\!\delta^{2}G^{\alpha\beta}[h^{(1)},h^{(1)}]\nabla_{\beta}\varphi_{\alpha}dV \\
    = \int_{\rho=s} \!\!\delta^{2}G^{\alpha\beta}[h^{(1)},h^{(1)}]\varphi_{\alpha}dS_\beta.
\end{multline}

We evaluate the surface integral in Eq.~\eqref{eq:Bianchi IBP} using the expansions~\eqref{eq:ddG decomp},~\eqref{eq:FWdS}, \eqref{eq:d2GSSexpansion}, \eqref{eq:QR near gamma}, and a Taylor series in $x^i$ for $\varphi_\alpha$, which we can write as
\begin{equation}\label{eq:Taylor phibeta}
    \varphi_\alpha = \varphi_\alpha|_\gamma + sn^\beta\partial_\beta \varphi_\alpha|_{\gamma} + \O(s^2).
\end{equation}
The Taylor series can also be written covariantly using $\partial_\beta\varphi_\alpha|_\gamma = \nabla_\beta\varphi_\alpha|_\gamma + \O(\e)$, where the $\O(\e)$ term is proportional to the acceleration.

Consider first the singular-singular terms. Appealing to Eqs~\eqref{eq:d2GSSexpansion}, \eqref{eq:FWdS}, and~\eqref{eq:Taylor phibeta}, we find
\begin{equation}
    \int_{\rho=s}\delta^2G^{\alpha\beta}[h^{\S(1)},h^{\S(1)}]\varphi_{\alpha}dS_\beta = -\frac{28\pi m^{2}}{3s}\int_{\gamma}\nabla_\alpha\varphi^{\alpha} d\tau,\label{eq:Bianchi SS term1}
\end{equation}
where even powers of $\rho$ have vanished due to their parity, we have suppressed acceleration terms and $\O(s)$ terms, and we have noted that $u^\beta\partial_\beta\varphi_\alpha=d\varphi_\alpha/d\tau$ integrates to zero because $\varphi_\alpha$ has compact support. The other singular-singular term in Eq.~\eqref{eq:Bianchi on test function} is given by
\begin{align}
    -8\pi\bigl\langle \nabla_\beta\varphi_\alpha,T^{\alpha\beta}_{\S\S}\bigr\rangle ={}& \frac{4\pi m^2}{3s}\int_\gamma(7g^{\alpha\beta}-2u^\alpha u^\beta)\varphi_{\alpha;\beta}d\tau \nonumber \\
        ={}& \frac{28\pi m^{2}}{3s}\int_{\gamma}\nabla_\alpha\varphi^{\alpha} d\tau,\label{eq:Bianchi SS term2}
\end{align}
where we again note the second term vanishes when going to the second line as it is a total time derivative of $\varphi_{\alpha}$  (up to acceleration terms). From Eqs.~\eqref{eq:Bianchi SS term1} and~\eqref{eq:Bianchi SS term2}, we see that the two singular-singular contributions cancel in Eq.~\eqref{eq:Bianchi on test function}.

Moving to the singular-regular terms, we find Eqs.~\eqref{eq:QR near gamma}, \eqref{eq:FWdS}, and \eqref{eq:Taylor phibeta} imply
\begin{align}
    \MoveEqLeft[2] \lim_{s\to0}\int_{\rho=s}2Q^{\alpha\beta}_{R}[h^{\S(1)}]\varphi_\beta dS_\alpha \nonumber \\
        ={}& -\frac{4\pi m}{3}\int_{\gamma}\bigl(V^{\alpha\beta\mu\nu}\varphi_{\alpha}\nabla_\beta h^{\R(1)}_{\mu\nu}\nonumber\\[-2pt]
        &\qquad\qquad\quad + W^{\alpha\beta\mu\nu}\nabla_\beta\varphi_{\alpha}h^{\R(1)}_{\mu\nu}\bigr)d\tau,\label{eq:Bianchi SR term1}
\end{align}
and the $T^{\alpha\beta}_Q$ term in Eq.~\eqref{eq:Bianchi on test function} evaluates to
\begin{align}
    8\pi\bigl\langle \nabla_\beta\varphi_\alpha,T^{\alpha\beta}_Q\bigr\rangle = -\frac{4\pi m}{3}\int_{\gamma}X^{\alpha\beta\mu\nu}\nabla_\beta\varphi_{\alpha}h^{\R(1)}_{\mu\nu} d\tau.\label{eq:Bianchi SR term2}
\end{align}
Here
\begin{align}
    V^{\alpha\beta\mu\nu} \coloneqq{}& P^{\mu\nu}u^{\alpha}u^\beta - 4P^{\alpha(\mu}u^{\nu)}u^\beta + 3P^{\alpha\beta}u^\mu u^\nu, \\
    W^{\alpha\beta\mu\nu} \coloneqq{}& P^{\alpha\beta}(P^{\mu\nu}+2u^\mu u^\nu) - (P^{\alpha(\mu}+2u^\alpha u^{(\mu})P^{\nu)\beta},
\end{align}
and $X^{\alpha\beta\mu\nu}$ is given in Eq.~\eqref{eq:X}.

Combining Eqs.~\eqref{eq:Bianchi SR term1} and Eq.~\eqref{eq:Bianchi SR term2} in Eq.~\eqref{eq:Bianchi on test function}, and integrating by parts to move any time derivatives from $\varphi_\mu$ to $h^{\R(1)}_{\mu\nu}$, we find that
\begin{multline}
    \bigl\langle\varphi_\alpha,\nabla_\beta\widehat{\delta^{2}G}{}^{\alpha\beta}\bigr\rangle
    = -4\pi m\int_\gamma \varphi_{\alpha}U^{\alpha\beta\gamma\mu}\nabla_\mu h^{\R(1)}_{\beta\gamma}d\tau, \label{eq:divd2G}
\end{multline}
where $U^{\alpha\beta\mu\nu}$ is given by Eq.~\eqref{eq:Udef}. Comparison of Eq.~\eqref{eq:divd2G} with  Eq.~\eqref{eq:divTtilde} shows that the Bianchi identity~\eqref{eq:Bianchi 2nd order hat} is satisfied, as desired.

We reiterate the main conclusion: our distributional promotion of $\delta^2G^{\alpha\beta}$ satisfies an appropriate Bianchi identity that enforces
\begin{align}
    \tilde{\nabla}_{\beta}\Teff^{\alpha\beta} = \mathcal{O}(\varepsilon^3).
\end{align}
This conservation equation is equivalent to $\tilde u^\beta\tilde\nabla_\beta\tilde u^\alpha = \O(\e^2)$, which in turn is equivalent to the first-order self-force equation 
\begin{equation}
u^\beta\nabla_\beta u^\alpha = -\frac{\e}{2}P^{\alpha\mu}\Bigl(2\nabla_{\beta}h^{\R(1)}_{\gamma\mu}-\nabla_\mu h^{\R(1)}_{\beta\gamma}\Bigr)u^\beta u^\gamma + \O(\e^2),
\end{equation}
known as the MiSaTaQuWa equation~\cite{Poisson:2011nh}. Extending this method to obtain the second-order self-force equation, $\tilde u^\beta\tilde\nabla_\beta\tilde u^\alpha = \O(\e^3)$, derived in Ref.~\cite{Pound:2017psq} from the field equations outside the body, would require the \emph{third}-order Bianchi identity.

Finally, we comment on uniqueness. One could freely modify our definition of $\widehat{\delta^2G}{}^{\alpha\beta}$ by adding any distribution supported only on $\gamma$ (meaning a linear combination of $\delta$ functions and derivatives of $\delta$ functions). This would preserve the pointwise value of $\delta^2G^{\alpha\beta}$ for $x\notin\gamma$. Doing so would change $\tilde T^{\alpha\beta}$ in Eq.~\eqref{eq:fullEFELorenz}, while keeping $\deltaG^{\alpha\beta}[h^{(2)}]$ unchanged (our fundamental requirement). But one reason to consider our definition canonical is the Bianchi identity it satisfies and the resulting naturalness of the stress-energy conservation. Any other definition would yield the same equation of motion for the particle, since it would only move $\delta$-function terms between the two terms in Eq.~\eqref{eq:ConservationCondition}, but the equation of motion would not be so easily interpretable as arising from a conservation equation.

\section{Blanchet and Damour's Hadamard regularization}
\label{sec:Hadamard}

In this section we show that our distributional definition of $\delta^2G^{\alpha\beta}$ in Eq.~\eqref{eq:ddGhat distribution} is equivalent to defining it via Hadamard regularization in Eq.~\eqref{eq:ddG = PfddG}. As explained in the Introduction, this implies that treating the source in either way yields the same physical solution for $h^{(2)}_{\mu\nu}$.

More concretely, we want to show that our distribution, which acts on test fields according to
\begin{align}
    \bigl\langle \varphi_{\alpha\beta},\widehat{\delta^2 G}{}^{\alpha\beta}\bigr\rangle &= \lim_{s\to0}\biggl\{\int_{\rho>s}\!\!\!\!\!\delta^2G^{\alpha\beta}[h^{(1)},h^{(1)}]\varphi_{\alpha\beta}dV\nonumber\\ 
     &\quad -8\pi\Bigl\langle\varphi_{\alpha\beta},T^{\alpha\beta}_{\S\S}\Bigr\rangle\biggr\} + 8\pi\Bigl\langle\varphi_{\alpha\beta},T^{\alpha\beta}_{Q}\Bigr\rangle,\label{eq:d2G def}
\end{align}
is equivalent to 
\begin{multline}\label{d2Ghat FP}
    \bigl\langle \varphi_{\alpha\beta},\widehat{\delta^2 G}{}^{\alpha\beta}\bigr\rangle = \FP_{B=0}\int \left(\frac{\rho}{\rho_0}\right)^{\!B} \delta^2 G_{\mu\nu}[h^{(1)},h^{(1)}]\varphi^{\mu\nu}dV\\
    + 8\pi\Bigl\langle\varphi_{\alpha\beta},T^{\alpha\beta}_{Q}\Bigr\rangle\,. 
\end{multline}
To do this, let us split the domain of integration for Eq.~\eqref{d2Ghat FP} in the following way:
\begin{align}
    &\FP_{B=0}\biggl\{\int_{\rho>s} \left(\frac{\rho}{\rho_0}\right)^{\!B} \delta^2 G_{\mu\nu}[h^{(1)},h^{(1)}]\varphi^{\mu\nu}dV\nonumber
    \\
    &\quad +\int_{\rho<s} \left(\frac{\rho}{\rho_0}\right)^{\!B} \delta^2 G_{\mu\nu}[h^{(1)},h^{(1)}]\varphi^{\mu\nu}dV \biggr\} \nonumber
    \\
    &= \int_{\rho>s} \delta^2 G_{\mu\nu}[h^{(1)},h^{(1)}]\varphi^{\mu\nu}dV\nonumber
    \\
    &\quad +\FP_{B=0}\int_{\rho<s} \left(\frac{\rho}{\rho_0}\right)^{\!B} \delta^2 G_{\mu\nu}[h^{(1)},h^{(1)}]\varphi^{\mu\nu}dV. \label{eq:FPtoCounter}
\end{align}
Here $s>0$ is small, but we do not yet take the limit. In the second equality we use the fact that the integral for $\rho>s$ is perfectly regular even for $B=0$, meaning the finite part acts trivially.

Within the second integral in Eq.~\eqref{eq:FPtoCounter}, we use the decomposition~\eqref{eq:ddG decomp} into singular-singular, singular-regular, and regular-regular pieces of $\delta^2G^{\alpha \beta}$. The integral of $\delta^2G^{\alpha \beta}[h^{\R(1)},h^{\R(1)}]\varphi_{\alpha\beta}$ can be neglected as it contributes $\O(s^3)$. The integral of the singular-regular piece can also be neglected. To see this, use Eq.~\eqref{eq:QR near gamma}, from which it follows that
\begin{align}
    \FP_{B=0}\int_{\rho<s}\left(\frac{\rho}{\rho_0}\right)^{\!B} Q_\R^{\alpha\beta}[h^{\S(1)}]\varphi_{\alpha\beta}dV
    &=\mathcal{O}(s),
\end{align}
where we used the fact that the $\rho$ integral is finite due to the finite part but the integral over the angles for the $1/\rho^3$--piece of Eq.~\eqref{eq:QR near gamma} vanishes.

Finally, for the singular-singular term, we expand the integrand for small $\rho$. We use the local expansion~\eqref{eq:d2GSSexpansion}, combined with the Taylor series
\begin{equation}
    \varphi_{\alpha \beta}(\tau,\rho,\theta^A) =  \varphi_{\alpha \beta}|_\gamma + \rho n^\mu \partial_\mu \varphi_{\alpha \beta}|_\gamma + \mathcal{O}(\rho^2) \label{eq:varphiexpansion}
\end{equation}    
and with Eq.~\eqref{eq:FWdV}. 
Together they imply
\begin{align}
    &\int_{\rho<s} \left(\frac{\rho}{\rho_0}\right)^{\!B} \delta^2 G^{\alpha \beta}[h^{\S(1)},h^{\S(1)}]\varphi_{\alpha \beta}dV  \nonumber
    \\
   &= \int_{\rho<s} \left(\frac{\rho}{\rho_0}\right)^{\!B} \biggl[ \frac{m^2}{3\rho^2}(2u^\alpha u^\beta + 42\hat{n}^{\alpha\beta}-7\eta^{\alpha\beta})\varphi_{\alpha \beta}|_\gamma \nonumber
   \\
    &\qquad\qquad\qquad\qquad  + \mathcal{O}(\rho^0)\biggr] d\rho d\Omega d\tau.\label{eq:FP derivation step 2}
\end{align} 
$\partial_\alpha \varphi^{\mu \nu}|_\gamma$ naively enters in a $1/\rho$ term within the square brackets, but that term vanishes upon integrating over the sphere because it has odd parity. We evaluate the finite part of this using
\begin{equation}
    \FP_{B=0}\int_0^s \left(\frac{\rho}{\rho_0}\right)^{\!B} \frac{d\rho}{\rho^2} = \FP_{B=0}\frac{s^{B-1}}{\rho^B_0(B-1)} = -\frac{1}{s},
\end{equation}
which leads to
\begin{align}    
    &\FP_{B=0}\int_{\rho<s} \left(\frac{\rho}{\rho_0}\right)^{\!B} \delta^2 G^{\alpha\beta}[h^{(1)},h^{(1)}]\varphi_{\alpha\beta}dV \nonumber\\
    &= \frac{4\pi m^2}{3s}\int (7 g^{\alpha\beta}-2 u^\alpha u^\beta)\varphi_{\alpha\beta}d\tau + \mathcal{O}(s).
\end{align}
This, up to order $s$, is simply $-8\pi\bigl\langle \varphi_{\alpha\beta},T^{\alpha\beta}_{\S\S}\bigr\rangle$, as we see from Eq.~\eqref{eq:Tcount}.

Since the above calculations apply for any small $s>0$, we can freely take the limit $s\to0$ in Eq.~\eqref{eq:FPtoCounter}. Using our results for each of the integrals, we then obtain
\begin{align}
    &\FP_{B=0}\int \left(\frac{\rho}{\rho_0}\right)^{\!B} \delta^2 G^{\alpha\beta}[h^{(1)},h^{(1)}]\varphi_{\alpha\beta}dV\nonumber\\
    &= \lim_{s\to0} \biggl\{\int_{\rho>s} \delta^2 G^{\alpha\beta}[h^{(1)},h^{(1)}]\varphi_{\alpha\beta}dV - 8\pi\Bigl\langle \varphi_{\alpha\beta},T^{\alpha\beta}_{\S\S}\Bigr\rangle \biggr\}.
\end{align}
This establishes our claim: $\widehat{\delta^2G}{}^{\alpha\beta}$ can be represented using a Hadamard finite part as in Eq.~\eqref{d2Ghat FP}, which is distributionally equivalent to Eq.~\eqref{eq:d2G def}.

\section{Applications}
\label{sec:applications}

In this section we focus on the practical implementation of the distributional second-order equations. We first show how solutions can be obtained using Green's functions. We then reformulate our regularization method for practical implementation in (spherical or spheroidal) mode calculations. Then we discuss the implementation for scalar equations obtained from the Einstein equation, with the Teukolsky equation as the prime example. Due to the extra derivatives in these scalar equations, the singularities in the source become one or two degrees worse. We end with a demonstration of our methods in a toy example for the radial Teukolsky equation. 

\subsection{Green's function representations of solution}

One advantage of a distributional field equation is that it allows us to write the retarded solution $h^{(2)}_{\alpha\beta}$ as an integral against a retarded Green's function. 

There does not exist a Green's function for the complete linearised Einstein operator $\deltaG^{\alpha\beta}$ due to its invariance under a linear gauge transformation; one must first specify the gauge. Let us specialize to the Lorenz gauge, in which case the second-order field equations are 
\begin{equation}
    \calE^{\mu\nu}[\barh^{(2)}] = 8\pi \tilde T_{(2)}^{\mu\nu} -\widehat{\delta^{2}G}{}^{\mu\nu}[h^{(1)},h^{(1)}]\,. \label{eq:EFE}
\end{equation}
Here 
\begin{equation}
\calE^{\mu\nu}[\bar h] \coloneqq -\frac{1}{2}\left(\Box \bar h^{\mu\nu} +2 R^\mu{}_\alpha{}^\nu{}_\beta\bar h^{\alpha\beta}\right)
\end{equation}
denotes $\deltaG^{\mu \nu}[h]$ in Lorenz gauge, and $\bar h^{(2)}_{\mu \nu}\coloneqq h^{(2)}_{\mu \nu}-\frac{1}{2}g_{\mu \nu} g^{\alpha\beta}h^{(2)}_{\alpha\beta}$ is the trace-reversed metric perturbation. 

Let us denote the retarded Green's function for $\calE_{\mu\nu}$ by \(G_{\mu\nu\mu'\nu'}(x,x')\). We then claim that the retarded solution to Eq.~\eqref{eq:EFE} is given by
\begin{align}
    \barh^{(2)}_{\mu\nu}(x) &= \bigl\langle G_{\mu\nu\mu'\nu'}(x,x'), 8\pi \tilde T_{(2)}^{\mu'\nu'} -\widehat{\delta^{2}G}{}^{\mu'\nu'}\bigr\rangle_{x'}\\
    &= 8\pi\int G_{\mu\nu\mu'\nu'}(x,x')(\tilde T_{(2)}^{\mu'\nu'}-T_Q^{\mu'\nu'})dV' \nonumber \\
        &\quad -\lim_{s\to 0}\biggl\{\int_{\rho'>s}\!\!\!\!G_{\mu\nu\mu'\nu'}\!(x,x')\delta^{2}G^{\mu'\nu'}[h^{(1)},h^{(1)}]dV'  \nonumber
        \\
         &\qquad \qquad - {8\pi} \int G_{\mu\nu\mu'\nu'}(x,x')T^{\mu' \nu'}_{\SS}dV'\biggr\} \label{eq:h^2Lor}
\end{align}
with $T^{\mu' \nu'}_{\SS}$ and $T^{\mu' \nu'}_Q$ given in Eqs.~\eqref{eq:Tcount} and~\eqref{eq:TQ}. Since the Green's function, unlike a test function, does not have compact support, we might question whether Eq.~\eqref{eq:h^2Lor} is actually  a solution to Eq.~\eqref{eq:EFE}. Establishing this requires showing that $\mathcal{E}_{\mu \nu}[\bar h^{(2)}]$ is the same distribution as the source on the right-hand side. 

If Eq.~\eqref{eq:h^2Lor} did not involve the limit $s\to0$, we would immediately obtain the desired result using the fact that the Green's function is a fundamental solution, satisfying
\begin{multline}\label{eq:EG=delta}
-\frac{1}{2}\bigl[\Box G_{\mu\nu}{}^{\mu'\nu'}+2R_{\mu}{}^\alpha{}_\nu{}^\beta G_{\alpha\beta}{}^{\mu'\nu'}\bigr] \\= \delta^{\mu'}_{(\mu} \delta^{\nu'}_{\nu)} \delta^4(x,x').
\end{multline}
But the presence of the limit makes the analysis more subtle. To make it clear, we examine the action of $\calE^{\mu\nu}[h^{(2)}]$ on a test field $\varphi_{\mu\nu}$. We first recall
\begin{equation}\label{eq:<phi,Eh2>}
    \bigl\langle \varphi_{\mu\nu} , \mathcal{E}^{\mu\nu}[ h^{(2)}]\bigr\rangle \coloneqq \bigl\langle \mathcal{E}^{\mu\nu}[\varphi],  h^{(2)}_{\mu\nu}\bigr\rangle
\end{equation}
since $\mathcal{E}^{\mu\nu}$ is self-adjoint. Noting that the right-hand side of Eq.~\eqref{eq:<phi,Eh2>} is an ordinary integral, we can write $\bar h^{(2)}_{\mu\nu}=\lim_{s\to0}\bar h^{s(2)}_{\mu\nu}$ and appeal to the dominated convergence theorem to move the limit outside the integral:
\begin{equation}\label{eq:<phi,Eh2>=lim<Ephi,hs2>}
    \bigl\langle \varphi_{\mu\nu} , \mathcal{E}^{\mu\nu}[ h^{(2)}]\bigr\rangle = \lim_{s\to0}\bigl\langle \mathcal{E}^{\mu\nu}[\varphi],  h^{s(2)}_{\mu\nu}\bigr\rangle.
\end{equation}
Here $\bar h^{s(2)}_{\mu\nu}$ is the entire right-hand side of Eq.~\eqref{eq:h^2Lor} with the limit $s\to0$ omitted.

Now we can appeal to 
\begin{align}
\Bigl\langle \calE^{\mu\nu}[\varphi]&, \int G_{\mu\nu\mu'\nu'}S^{\mu'\nu'}dV'\Bigr\rangle \nonumber\\
&= \int\bigl\langle \calE^{\mu\nu}[\varphi],  G_{\mu\nu\mu'\nu'}\bigr\rangle S^{\mu'\nu'}dV'\\
&= \int\varphi_{\mu'\nu'} S^{\mu'\nu'}dV',\label{eq:<Ephi,int GS>}
\end{align}
for any integrable $S^{\mu'\nu'}$, where we interchanged integrals in the first equality and used Eq.~\eqref{eq:EG=delta} in the second. Substituting Eq.~\eqref{eq:h^2Lor} in Eq.~\eqref{eq:<phi,Eh2>=lim<Ephi,hs2>} and using Eq.~\eqref{eq:<Ephi,int GS>} yields our desired result:
    \begin{align}
    \bigl\langle \varphi_{\mu\nu} , \mathcal{E}^{\mu\nu}[\bar h^{(2)}]\bigr\rangle &=  \bigl\langle \varphi_{\mu\nu} , 8\pi\bigl(\tilde T^{\mu\nu}_{(2)}-T^{\mu\nu}_Q\bigr)\bigr\rangle\nonumber\\ 
    &\quad - \lim_{s\to0}\bigl\langle\varphi_{\mu\nu}, \theta(\rho-s)\delta^2G^{\mu\nu}-8\pi T^{\mu\nu}_{\S\S}\bigr\rangle\nonumber\\
    &=\bigl\langle\varphi_{\mu\nu} , 8 \pi \tilde T^{\mu\nu}_{(2)}- \widehat{\delta^2 G}{}^{\mu\nu} \rangle. \label{eq:proofofGreens}
\end{align}

In the above derivation we used the counterterm form of $\widehat{\delta^2G}{}^{\mu\nu}$. Equivalently, we can use its Hadamard form~\eqref{eq:ddG = PfddG}. In that case the appropriate Green's-function representation of $\bar h^{(2)}_{\mu\nu}$ becomes
\begin{align}
    \barh^{(2)}_{\mu\nu} &= 8\pi\int G_{\mu\nu\mu'\nu'}(\tilde T_{(2)}^{\mu'\nu'}-T_Q^{\mu'\nu'})dV'\nonumber 
    \\
    &\quad -\int_{\rho'>a}  G_{\mu\nu\mu'\nu'}\delta^{2}G^{\mu'\nu'}[h^{(1)},h^{(1)}]dV' \nonumber
    \\
    &\quad -\FP_{B=0}\int_{\rho'<a} \biggl(\frac{\rho'}{\rho_0}\biggr)^{\!B} G_{\mu\nu\mu'\nu'}\delta^{2}G^{\mu'\nu'}[h^{(1)},h^{(1)}]dV'\,, \label{eq:GreensonSol}
\end{align}
where $a>0$ is an arbitrary cutoff. The cutoff is necessary since the Green's function, unlike a test function, does not have compact support. However, the end result will be independent of the cutoff. To see this, take the derivative with respect to $a$ of Eq. \eqref{eq:GreensonSol}:
\begin{align}
    \partial_a  \barh^{(2)}_{\mu\nu} 
    &= \int_{\rho'=a}  G_{\mu\nu\mu'\nu'}\delta^{2}G^{\mu'\nu'}[h^{(1)},h^{(1)}]dS' \nonumber
    \\
    &\ -\FP_{B=0}\int_{\rho'=a}\! \biggl(\frac{a}{\rho_0}\biggr)^{\!B} \! \! G_{\mu\nu\mu'\nu'}\delta^{2}G^{\mu'\nu'}[h^{(1)},h^{(1)}]dS'.
\end{align}
Since $a>0$, the second integral is finite for $B=0$, the finite part operation acts trivially, and the two terms above cancel.

To show that Eq.~\eqref{eq:GreensonSol} is equivalent to Eq.~\eqref{eq:h^2Lor}, using the fact that the result is independent of $a$, we take the limit $a \to 0$:
\begin{align}
    \barh^{(2)}_{\mu\nu} &= 8\pi\int G_{\mu\nu\mu'\nu'} (\tilde T_{(2)}^{\mu'\nu'}-T_Q^{\mu'\nu'})dV' \nonumber \\
        &\quad - \lim_{a\to 0}\biggl\{\int_{\rho'>a}G_{\mu\nu\mu'\nu'}\delta^{2}G^{\mu'\nu'} [h^{(1)},h^{(1)}]dV' \nonumber \\
        &\quad + \FP_{B=0}\int_{\rho'<a}\biggl(\frac{\rho'}{\rho_0}\biggr)^{\!B} G_{\mu\nu\mu'\nu'}\delta^{2}G^{\mu'\nu'}[h^{(1)},h^{(1)}]dV'\biggr\}.
\end{align}
We then expand the last term around the worldline in Fermi-Walker coordinates and evaluate the finite part to find 
\begin{align}
    \MoveEqLeft[2] \FP_{B=0}\int_{\rho'<a}\Bigl(\frac{\rho'}{\rho_0}\Bigr)^{\!B} G_{\mu\nu\mu'\nu'}\delta^{2}G^{\mu'\nu'}[h^{(1)},h^{(1)}]{\rho'}^2d\tau'd\rho'd\Omega' \nonumber \\
            ={}& \frac{4\pi m^2}{3a}\int G_{\mu\nu\mu'\nu'}[7g^{\mu'\nu'} - 2u^{\mu'}u^{\nu'} + \mathcal{O}(a^2)]d\tau'.
\end{align}
Exchanging $a$ for $s$ we see that we get back exactly Eq.~\eqref{eq:h^2Lor}, establishing that the Hadamard regularization is equivalent to the counterterm at the level of Green's-function representations.

\subsection{Mode calculations in black hole spacetimes}

So far, the regularization procedure has been formulated using the proper distance from the worldline, $\rho$. This is a natural variable to work with in Fermi-Walker coordinates, but it is generally not well suited to practical second-order computations in black hole spacetimes, where one almost exclusively works in coordinates centered on the primary black hole. In addition, one would normally want to carry out calculations at the level of modes~\cite{Martel:2005ir,Pound:2021qin,Miller:2023ers,Spiers:2023mor,Upton:2025bja}. In the following, we show how the regularization scheme can be straightforwardly reformulated to meet these specifications.

To obtain a useful representation of our distributional source at the level of modes, we focus on its Hadamard form~\eqref{eq:ddG = PfddG}. Since the Einstein equations do not separate under a mode decomposition in Kerr spacetime, we restrict to a Schwarzschild background. Working in Schwarzschild coordinates $x^\alpha=(t,r,\theta,\phi)$ and a complex null tetrad $e^\alpha_a = (l^\alpha,n^\alpha,m^\alpha,\bar m^\alpha)$, we expand the quadratic source term in spin-weighted spherical harmonics ${}_sY_{\ell m}$:
\begin{align}
    \delta^2 G^{ab}[h^{(1)},h^{(1)}] = \sum_{\ell=|s|}^\infty \sum_{m=-\ell}^\ell \delta^2 G^{ab}_{\ell m}(t,r)\; {}_sY_{\ell m} (\theta, \phi)\, ,\label{eq:ddG mode sum}
\end{align}
where the spin weight is $0$, $\pm1$, or $\pm2$ depending on the tetrad component.

We then have
\begin{align}
   \bigl\langle \varphi_{\alpha\beta}&,{\rm Pf}\,\delta^2G^{\alpha\beta}\bigr\rangle \nonumber
   \\
   &=\FP_{B=0}\,\lim_{\ell_{\text{max}\to \infty}}\int \biggl(\frac{\rho}{\rho_0}\biggr)^{\!B}\varphi_{ab}\,\delta^2 G^{ab}_{\ell_{\rm max}}\, dV\,, 
\end{align}
where $\delta^2 G^{ab}_{\ell_{\rm max}}$ denotes the mode sum~\eqref{eq:ddG mode sum} truncated at $\ell=\ell_{\text{max}}$, and we note
\begin{multline}\label{eq:contraction in tetrad}
    \varphi_{ab}\delta^2G^{ab} = \varphi_{ll}\delta^2G^{nn} + \varphi_{ln}\delta^2G^{nl} -\varphi_{lm}\delta^2G^{n\bar m} \\
    -\varphi_{l\bar m}\delta^2G^{nm}+\ldots +\varphi_{\bar m\bar m}\delta^2G^{mm}. 
\end{multline}
Since the integral is evaluated in a region of $B$ where it is finite, we can insert an additional, one-dimensional finite-part operation: 
\begin{align}    
    \bigl\langle& \varphi_{\alpha\beta},{\rm Pf}\,\delta^2G^{\alpha\beta}\bigr\rangle \nonumber\\
    &= \FP_{B=0}\,\lim_{\ell_{\text{max}\to \infty}}\FP_{B'=0}\int |\Delta \tilde{r}|^{B'}\biggl(\frac{\rho}{\rho_0}\biggr)^{\!B} \varphi_{ab}\,\delta^2 G^{ab}_{\ell_{\text{max}}}\,dV\,, 
\end{align}
where $\Delta \tilde r \coloneqq \frac{r-r_p}{r_p}$. The outermost finite part now does no work; the expression is finite at $B=0$. Hence, dropping it, we obtain 
\begin{align}    
    \bigl\langle \varphi_{\alpha\beta}&,{\rm Pf}\,\delta^2G^{\alpha\beta}\bigr\rangle \nonumber\\
    &=\lim_{\ell_{\text{max}}\to\infty}\FP_{B'=0}\int |\Delta \tilde r|^{ B'}  \varphi_{ab}\, \delta^2 G^{ab}_{\ell_{\text{max}}} dV\,.\label{eq:<phi,PfddG> modes}
\end{align}
We caution the reader that the order of operations is essential here. If the limit $\ell_{\text{max}}\to\infty$ were taken before the finite part, the result would diverge because the factor $|\Delta \tilde r|^{ B'}$ does not regularize the integral over angles.

Using $dV=r^2dtdrd\Omega$, one finds the integration over angles reduces to
\begin{equation}
    \int \varphi_{ab}\, {}_s Y_{\ell m} d\Omega = (\varphi^{*\, \ell m}_{ab})^*.
\end{equation}
Here we use $\varphi_{ab}=\sum_{\ell m}\varphi^{\ell m}_{ab}\,{}_{-s}Y_{\ell m}$, noting that $\varphi_{ab}$ always comes with the opposite spin weight than $\delta^2G^{ab}$ in Eq.~\eqref{eq:contraction in tetrad}. We also used ${}_sY_{\ell m} = (-1)^{m+s} {}_{-s}Y^*_{\ell, -m}$, and $(-1)^{m+s}(\varphi^{\ell,- m}_{ab})^*=\varphi^{*\, \ell m}_{ab}$~\cite{Spiers:2023mor}, where $\varphi^{*\ell m}_{ab}$ denote the mode coefficients of the complex conjugate $\varphi^*_{ab}$ (noting that components involving $m^\alpha$ or $\bar m^\alpha$ are complex if $\varphi_{\alpha\beta}$ is real). This reduces Eq.~\eqref{eq:<phi,PfddG> modes} to a two-dimensional integral,
\begin{align}
   &\bigl\langle\varphi_{ab}, {\rm Pf}\,\delta^2G^{ab}\bigr\rangle \nonumber\\
   &= 
   \sum_{\ell,m} \FP_{B=0} \int\! dt\, dr\, r^2 |\Delta \tilde{r}|^B \delta^2 G^{ab}_{\ell m}\, (\varphi^{*\,\ell m}_{ab})^*\,.
\end{align}

When working at the level of modes, one can skip directly to the two-dimensional action of a distribution on a two-dimensional test field $\varphi_{ab}(t,r)$,
\begin{align}
    \bigl\langle \varphi_{ab}, {\rm Pf}\delta^2 G^{ab}_{\ell m}\bigr\rangle_{2\text{D}}  \coloneqq  \FP_{B=0} \int\! dt\, dr\,  |\Delta \tilde{r}|^B \delta^2 G^{ab}_{\ell m}\, \varphi_{ab}(t,r).
\end{align} 
We can straightforwardly evaluate integrals of this form. In generic second-order self-force calculations, the singular behaviour of the mode-decomposed source is expected to be~\cite{Upton:2025bja} 
\begin{equation}\label{eq:ddGlm singularity}
\delta^2 G^{ab}_{\ell m} = \frac{A^{ab}_{\ell m}}{\Delta r^2} + \frac{B^{ab}_{\ell m}}{\Delta r} + \mathcal{O}(\log|\Delta r|) 
\end{equation}
with $\Delta r := r-r_p$ and some functions $A^{ab}_{\ell m}(t)$ and $B^{ab}_{\ell m}(t)$. From this we see that
\begingroup\allowdisplaybreaks%
\begin{align}\label{eq:FP example}
    &\bigl\langle \varphi_{ab}, {\rm{Pf}}\,\delta^2 G^{ab}_{\ell m} \bigr\rangle_{2\text{D}} \nonumber
    \\*
    &= \FP_{B=0} \int dt\, dr \left( \delta^2 G^{ab}_{\ell m} -\frac{A^{ab}_{\ell m}}{\Delta r^2} - \frac{B^{ab}_{\ell m}}{\Delta r} \right) |\Delta \tilde{r}|^B \varphi_{ab}\nonumber 
    \\*
    &\quad+ \FP_{B=0} \int dt\, dr  \left( \frac{A^{ab}_{\ell m}}{\Delta r^2} + \frac{B^{ab}_{\ell m}}{\Delta r} \right) |\Delta \tilde{r}|^B\varphi_{ab}  \nonumber
    \\
    &= \int dt\, dr \left( \delta^2 G^{ab}_{\ell m} -\frac{A^{ab}_{\ell m}}{\Delta r^2} - \frac{B^{ab}_{\ell m}}{\Delta r} \right)\varphi_{ab} \nonumber
    \\*
    & \quad- \int dt\, dr\,  A^{ab}_{\ell m} \log(|\Delta r|/r_p) \partial_r^2 \varphi_{ab}  \nonumber
    \\*
    &\quad-  \int dt\,dr\,    B^{ab}_{\ell m} \log(|\Delta r|/r_p) \partial_r\varphi_{ab}\,.
\end{align}\endgroup
In the first equality we added zero by adding and subtracting the regularized integral of the most singular terms. In the second equality, we repeatedly integrated by parts until we could safely move the $\FP$ inside the integral. For example,
\begin{align}
    \FP_{B=0}\int \frac{|\Delta \tilde r|^B}{\Delta r^2}\varphi_{ab}dr &= \FP_{B=0}\int_{\Delta r>0} \frac{\Delta r^{B-2}}{r_p^B}\varphi_{ab}dr \nonumber\\
    &\quad + \FP_{B=0}\int_{\Delta r<0} \frac{(-1)^B\Delta r^{B-2}}{r_p^B}\varphi_{ab}dr\nonumber\\
    &= \FP_{B=0}\int \frac{|\Delta r|^{B}}{r_p^BB(B-1)}\partial_r^2\varphi_{ab}dr \nonumber\\
    &= -\int [1+\log(|\Delta r|/r_p)]\partial_r^2\varphi_{ab}dr.\label{eq:FPexplicit}
\end{align}
Since the test field has compact support, an integral of the form ${\int \partial_r^2\varphi_{ab}dr}$ vanishes, meaning the first term in the parentheses vanishes and the length scale $r_p$ in the logarithm can be replaced with any other length scale (e.g., $M$). 

Finally, we note that equation~\eqref{eq:FP example} can equivalently be written in the following way using a principle value integral (around $r=r_p$):
\begin{align}
     &\bigl\langle \varphi_{ab}, {\rm{Pf}}\,\delta^2 G^{ab}_{\ell m} \bigr\rangle_{2\text{D}} \nonumber
    \\*
    &=\text{P.V.} \int dt\, dr \left( \delta^2 G^{ab}_{\ell m} -\frac{A^{ab}_{\ell m}}{\Delta r^2} \right)\varphi_{ab} \nonumber
    \\*
    & \quad- \int dt\, dr\,  A^{ab}_{\ell m} \log(|\Delta r|/r_p) \partial_r^2 \varphi_{ab}\,,
\end{align}
which allows one to avoid calculating $B_{\ell m}$.  

We can also write a Green's-function representation at the level of modes. The equivalent of \eqref{eq:GreensonSol} for the mode-decomposed metric perturbation, $h^{(2,\ell m)}_{ab}$, is given by
\begin{align}
     &\bar h^{(2,\ell m)}_{ab}(t,r)\nonumber\\
     &= 8\pi\int G^{\ell m}_{aba'b'}(t-t',r,r') (\tilde T_{(2,\ell m)}^{a'b'}-T_{Q\,\ell m}^{a'b'}) dt'dr'\nonumber 
     \\
    &\quad -\FP_{B=0}{\int_{|\Delta r|<a}} |\Delta \tilde{r}'|^B  G^{\ell m}_{aba'b'}(t-t',r,r') \delta^2 G^{a'b'}_{\ell m} dt'dr' \nonumber
    \\
    &\quad -{\int_{|\Delta r|>a}} G^{\ell m}_{aba'b'}(t-t',r,r') \delta^2 G^{a'b'}_{\ell m} dt'dr', \label{eq:hlm Greens function rep}
\end{align}
where $G_{aba'b'}^{\ell m}(t-t',r,r')$ is the Green's function for the mode-decomposed field equations.

\subsection{Teukolsky variables and other master scalars}

In black hole perturbation theory, rather than working with metric perturbations, it is often useful to solve scalar field equations~\cite{Martel:2005ir,Pound:2021qin,Spiers:2023mor,Spiers:2023cip} such as the Teukolsky equation \cite{Teukolsky:1972my,Teukolsky:1973ha}, the Sasaki-Nakamura equation~\cite{Sasaki:1981kj}, or the Regge-Wheeler and Zerilli equations~\cite{Regge:1957td,Zerilli:1970se}. These equations are simpler to solve than the full set of Einstein's field equations, and the master scalars one obtains from them capture the radiating degrees of freedom. Moreover, at second order, these scalar variables are invariant under linear gauge transformations of $h^{(2)}_{\alpha\beta}$ (though not under transformations of $h^{(1)}_{\alpha\beta}$)~\cite{Campanelli:1998jv,Spiers:2023cip}. Metric reconstruction techniques~\cite{Wald:1978vm,Kegeles:1979an,Chrzanowski:1975wv} allow one to obtain most of the metric perturbation directly from the scalars~\cite{Green:2019nam,Spiers:2023cip, Li:2026rkf}.

A second-order Teukolsky formalism in the self-force context was recently implemented, for the first time, in Ref.~\cite{Leather:2026zhl}. In that work, two of us utilized a puncture scheme. However, as pointed out in Ref.~\cite{Spiers:2023cip}, there are distinct advantages to working with distributional Teukolsky equations rather than punctured ones.

Here, rather than specializing to a particular scalar equation, we assume a generic second-order scalar equation of the form
\begin{align}\label{eq:Opsi2=S2}
    \hat{\mathcal{O}} \psi^{(2)} = S^{(2)}\,,
\end{align}
where $\hat{\mathcal{O}}$ is a second-order linear differential operator, $\psi^{(2)}$ is the scalar of interest, and  the source $S^{(2)}$  is given by\footnote{Here we work with lowered indices on the Einstein source so that expressions for $\hat{\mathcal{S}}^{\mu \nu}$ can easily be read off in familiar form from Refs.~\cite{Pound:2021qin,Spiers:2023mor}, for example.}
\begin{align}
    S^{(2)} = \hat{\mathcal{S}}^{\mu \nu}\bigl( 8\pi  \tilde T_{\mu \nu}^{(2)} - \widehat{\delta^2 G}{}_{\mu \nu}\bigr)\,.
\end{align} 
Here $\hat{\mathcal{S}}^{\mu \nu}$ is another linear differential operator acting on the source from the Einstein equation. Since the source in the Einstein equation is well defined as a distribution, it follows that $S^{(2)}$ is also well defined as
\begin{align}
    \langle \varphi, S^{(2)}\rangle &= 8\pi\int \left( \tilde T_{\mu \nu}^{(2)} - T^Q_{\mu \nu}\right) \hat{\mathcal{S}}^{\dagger\mu\nu}\varphi\, dV \nonumber
    \\
    &\quad -\FP_{B=0}\int     \left(\frac{\rho}{\rho_0}\right)^B \delta^2 G{}_{\mu \nu}\, \hat{\mathcal{S}}^{\dagger\mu\nu}\varphi\, dV.
\end{align}

In practical calculations we are again interested in working with Green's functions. This is done in a similar way as in Eq.~\eqref{eq:GreensonSol}: the retarded field is given by
\begin{align}
    \psi^{(2)}  &= 8\pi\int G(x,x')\hat{\mathcal{S}}^{\mu'\nu'} \bigl(\tilde T_{\mu' \nu'}^{(2)} - T^Q_{\mu' \nu'}\bigr) dV' \nonumber
    \\
    &\quad -\FP_{B=0}\int_{\rho'<a}     G(x,x')\, \hat{\mathcal{S}}^{\mu'\nu'}\!\biggl[ \biggl(\frac{\rho'}{\rho_0}\biggr)^{\!B} \delta^2 G_{\mu' \nu'}\biggr] \, dV' \nonumber
    \\
    &\quad -\int_{\rho'>a}    G(x,x')\, \hat{\mathcal{S}}^{\mu'\nu'}\delta^2 G_{\mu' \nu'} \, dV', \label{eq:solTeukolsky}
\end{align}
where $G(x,x')$ is the retarded Green's function for $\hat{\mathcal{O}}$ and $a>0$ is an arbitrary cut-off. First, we note that again the expression is independent of the cut-off $a$. To see this we take the derivative with respect to $a$:
\begin{align}
    \partial_a\psi^{(2)}  &=- \FP_{B=0}\int_{\rho'=a}     G(x,x')\, \hat{\mathcal{S}}^{\mu'\nu'}\!\left[ \left(\frac{\rho'}{\rho_0}\right)^{\!B} \delta^2 G_{\mu' \nu'}\right]\! dS' \nonumber
    \\
    &\quad +\int_{\rho'=a}    G(x,x')\, \hat{\mathcal{S}}^{\mu'\nu'}\delta^2 G_{\mu' \nu'} \, dS' \nonumber
    \\
    &= -\FP_{B=0}\int_{\rho'=a}     G(x,x')\, \hat{\mathcal{S}}^{\mu'\nu'}\!\left[ \left(\frac{\rho'}{\rho_0}\right)^{\!B} \delta^2 G_{\mu' \nu'}\right] \! dS' \nonumber
    \\
    &\quad +\FP_{B=0}\int_{\rho'=a} \left(\frac{\rho}{\rho_0}\right)^{\! B}   G(x,x')\, \hat{\mathcal{S}}^{\mu'\nu'}\delta^2 G_{\mu' \nu'} \, dS' \nonumber
    \\
    &=0\,, \label{eq:dapsi2}
\end{align}
where in the second equality we used that the integral on the second line is already finite, so we can trivially add a finite part operation on it. For the last equality we used that the difference between the two integrals will be proportional to derivatives of $(\rho/\rho_0)^B$, which will bring down a power of $B$. Since the rest of the integral is finite, the power of $B$ will ensure everything vanishes under the $\FP_{B=0}$ operation.

In the expression for $\psi^{(2)}$ one might also wonder about the placement of $(\rho'/\rho_0)^B$ relative to the $\hat{S}$ operator in the finite part integral. However, one can show that the placement does not matter:
\begin{align}
    &\FP_{B=0}\int_{\rho'<a}    G(x,x')\, \hat{\mathcal{S}}^{\mu'\nu'}\!\left[ \delta^2 G_{\mu' \nu'}  \left(\frac{\rho'}{\rho_0}\right)^{\!B} \right]\, dV' \nonumber
    \\
    &=\FP_{B=0}\int_{\rho'<a}     \left(\frac{\rho'}{\rho_0}\right)^{\!B}G(x,x')\, \hat{\mathcal{S}}^{\mu'\nu'}\delta^2 G_{\mu' \nu'} \, dV'. 
  \label{eq:FPdifference}
\end{align}
Establishing this equality looks as though it would be similar to the calculation in Eq.~\eqref{eq:dapsi2}, but the crucial difference here is that these integrals are not finite for $B=0$, so one has to be more careful. First, we note that $\hat{S}^{\mu \nu}$ is a smooth second-order differential operator. If we consider the \emph{difference} between the left- and right-hand sides of Eq.~\eqref{eq:FPdifference}, then zeroth derivatives cancel, and the form of $\hat{S}^{\mu \nu}$ implies all terms will be of the form
\begin{align}
    &\FP_{B=0} \int_{\rho<a} \left( f^{ij}(x,x') \frac{1}{\rho'^{N}}\partial_{ij}'\rho'^B+ g^{i}(x,x')\frac{1}{\rho'^{\tilde N}} \partial_{i}'\rho'^B \right)dV'
\end{align}
in Fermi-Walker coordinates. Here $N$ and $\tilde N$ are positive integers no larger than $5$ (and $\tilde N \geq2$)\footnote{If $\tilde N =1$ then the integral is already finite for $B=0$ and no integration by parts is needed. The integral then vanishes due to the factor of $B$ from the derivatives similar to what happens in Eq.~\eqref{eq:dapsi2}.}, given the divergence structure of $\delta^2 G_{\mu \nu}$. We then find that
\begin{align}
    &\FP_{B=0} \int_{\rho<a} \left( f^{ij} \frac{1}{\rho'^{N}}\partial_{ij}'\rho'^B+ g^{i}\frac{1}{\rho'^{\tilde N}} \partial_{i}'\rho'^B \right)dV' \nonumber
    \\
    &=\FP_{B=0} \int_{\rho'<a}\! \! \bigg( f^{ij}_V\frac{B[(B-2)n_{i}' n_j' +\delta_{ij}]}{\rho'^{N-B}} \nonumber\\
    &\qquad\qquad\qquad + g^{i}_Vn_{i}'\frac{B}{\rho'^{\tilde N-B-1}} \bigg) d\rho' d\Omega' d\tau' \nonumber
    \\
    &\cong\FP_{B=0} \int_{\rho'<a} \bigg(  \frac{\rho'^B (-1)^N\partial_\rho'^{N}\left(f^{ij}_V [(B-2)n_{i}' n_j'+\delta_{ij}]\right)}{(B-1)(B-2)\cdots (B+1-N)} \nonumber
    \\
    &\qquad\qquad+ \frac{\rho'^B (-1)^{\tilde N-1}\partial_\rho'^{\tilde N-1} (g^{i}_V n_{i}')}{(B-1)(B-2)\cdots (B+2-\tilde N)} \bigg) d\rho' d\Omega' d\tau'  \nonumber
    \\
    &=\int_{\rho'<a} \left(\frac{\partial_\rho'^{N}(f^{ij}_V [2n_{i}' n_j'-\delta_{ij}])}{(N-1)!} -\frac{\partial_\rho'^{\tilde N-1} (g^{i}_V n_{i}')}{(\tilde N -2)!}\right) \nonumber
    \\&\quad \quad d\rho' d\Omega' d\tau'\,. \label{eq:proofFPdif}
\end{align}
In the second line, we absorbed part of the volume element into the functions $f^{ij}$ and $g^i$, as in $f^{ij}(x,x')dV' = f^{ij}\rho'^2[1+\O(\e\rho',\rho'^2)] d\rho' d\Omega' d\tau' = f^{ij}_V\rho'^2d\rho' d\Omega' d\tau'$.  In the third line, we integrated by parts and used ``$\cong$'' to denote equality up to boundary terms; we have thrown away boundary terms as we showed that the final result cannot depend on $a$. Finally, we see that we are left with a total derivative in Eq. \eqref{eq:proofFPdif}, giving a pure boundary term we can again ignore. Hence, the difference between left- and right sides of Eq.~\eqref{eq:FPdifference} vanishes, establishing the equality.

In complete analogy with the proof in Eq.~\eqref{eq:proofofGreens}, one can show that Eq.~\eqref{eq:solTeukolsky} gives the right result, by which we mean
\begin{align}
    \langle \varphi, \hat{\mathcal{O}} \psi^{(2)} \rangle =\langle \varphi, S^{(2)} \rangle\,. \label{eq:psiGreens}
\end{align}
However, Eq.~\eqref{eq:psiGreens} is not enough. Equation~\eqref{eq:Opsi2=S2} is itself derived under the assumption that 
\begin{equation}\label{eq:psi2=Th2}
\psi^{(2)} = \hat{\mathcal{T}}^{\mu\nu} h^{(2)}_{\mu\nu}, 
\end{equation}
where $\hat{\mathcal{T}}^{\mu\nu}$ is a linear operator that (for smooth fields) acts on a metric perturbation to return the corresponding scalar~\cite{Spiers:2023cip}. Assuming Eq.~\eqref{eq:psi2=Th2}, one derives Eq.~\eqref{eq:Opsi2=S2} from Wald's operator identity~\cite{Wald:1978vm}
\begin{equation}\label{eq:Wald identity}
    \hat{\mathcal{O}} \hat{\mathcal{T}}^{\mu\nu}= \hat{\mathcal{S}}^{\alpha\beta} \hat{\delta G}_{\alpha\beta}{}^{\mu\nu},
\end{equation}
where we defined $\hat{\delta G}_{\alpha\beta}{}^{\mu\nu}h_{\mu\nu}\coloneqq \delta G_{\alpha\beta}[h]$.

Hence, for consistency, we must check that Eq.~\eqref{eq:psi2=Th2} holds with $\psi^{(2)}$ given by Eq.~\eqref{eq:solTeukolsky} and $h^{(2)}_{\mu\nu}$ by Eq.~\eqref{eq:GreensonSol}. We first observe that there does not exist a proper Green's function for the full linearised Einstein operator $\deltaG^{\mu\nu}$; $G_{\mu\nu\mu'\nu'}$ in Eq.~\eqref{eq:GreensonSol} is the Green's function for a gauge-fixed operator.\footnote{Even then, a true Green's function, in the sense of being a fundamental solution, only exists for certain gauge fixings. See Remark~4 on page~30 of Ref.~\cite{Casals:2024ynr}.} Without specifying which gauge we are working in, we can always write
\begin{align}
    h^{(2)}_{\mu\nu} = h^{(2)\text{Lor}}_{\mu\nu} + 2 \nabla_{(\mu} \xi_{\nu)}\,,
\end{align}
where $h^{(2)\text{Lor}}_{\mu\nu}$ is the Lorenz-gauge metric perturbation and $\xi^\mu$ is the gauge vector that transforms from the Lorenz gauge to the gauge of choice. Next, we note that $\hat{\mathcal{T}}^{\mu\nu}(\nabla_{(\mu} \xi_{\nu)}) =0$, even for distributional $\xi_{\nu}$ (a statement of linear gauge invariance, which holds for all scalars we consider). So, if we can show that $\psi^{(2)} = \hat{\mathcal{T}}^{\mu\nu} h^{(2)\text{Lor}}_{\mu\nu}$ then it will also hold for any gauge of choice.

For concreteness, we work with Teukolsky operators and let $G_{\mu\nu\mu'\nu'}$ denote the Lorenz-gauge Green's function. We then see that $\psi^{(2)} = \hat{\mathcal{T}}{}^{\mu\nu} h^{(2)\text{Lor}}_{\mu\nu}$, with $\psi^{(2)}$ given by Eq.~\eqref{eq:solTeukolsky} and $h^{(2)}_{\mu\nu}$ by Eq.~\eqref{eq:GreensonSol}, is equivalent to the following Green's function identity, which we prove below:
\begin{align}\label{eq:TG=GS}
    \hat{\mathcal{T}}{}^{\mu\nu} G_{\mu\nu\mu'\nu'}  = G(x,x') \hat{\mathcal{S}}_{\mu'\nu'} \,.
\end{align}
By the right-hand side, we mean an operator that acts on a non-compact test field, $f^{\mu' \nu'}$, (using $f$ instead of $\varphi$ here to signify that this is different from the usual compact test functions that we work with) at $x'$ as 
\begin{equation}
\langle f^{\mu'\nu'},G(x,x')\hat{\cal S}_{\mu'\nu'}\rangle_{x'} \coloneqq \langle \hat{\cal S}_{\mu'\nu'}f^{\mu'\nu'},G(x,x')\rangle_{x'}.
\end{equation}
We establish the identity~\eqref{eq:TG=GS} by showing that the two sides satisfy the same differential equation. Since they both satisfy the same boundary conditions, it follows that they are identical. To show they satisfy the same (distributional) differential equation, we act on them with the operator $\hat{\mathcal{O}}$. For the right-hand side, we see that 
\begin{align}\label{eq:OGS}
\hat{\mathcal{O}}G(x,x') \hat{\mathcal{S}}{}_{\mu'\nu'} = \delta^4(x,x')\hat{\mathcal{S}}_{\mu'\nu'},
\end{align}
which acts on a test function at $x'$ as
\begin{equation}\label{eq:<phi,OGS>}
    \langle f^{\mu'\nu'},\hat{\mathcal{O}}G(x,x')\hat{\cal S}_{\mu'\nu'}\rangle_{x'} = \hat{\cal S}_{\mu\nu}f^{\mu\nu}.
\end{equation}
For the left-hand side of Eq.~\eqref{eq:TG=GS}, we use the operator identity~\eqref{eq:Wald identity}, which implies
\begin{align}\label{eq:OTG}
\hat{\mathcal{O}} \hat{\mathcal{T}}^{\mu\nu} G_{\mu\nu\mu'\nu'}  &= \hat{\mathcal{S}}^{\alpha\beta}\hat{\deltaG}_{\alpha\beta}{}^{\mu\nu} G_{\mu\nu\mu'\nu'}.
\end{align}
Next, we note that 
\begin{align}
    \deltaG_{\alpha\beta}[h] =&\, -\frac{1}{2}(\Box \bar h_{\alpha\beta} + 2R_\alpha{}^\mu{}_\beta{}_\nu \bar h_{\mu\nu})\nonumber
    \\
    & + \nabla_{(\alpha}Z_{\beta)} -\frac{1}{2}g_{\alpha\beta}\nabla_\mu Z^\mu\,,
\end{align} 
where $Z^\alpha\coloneqq\nabla_\beta \bar h^{\alpha\beta}$. It follows that 
\begin{multline}\label{eq:dGG}
    \hat{\deltaG}_{\alpha\beta}{}^{\mu\nu}\bar G_{\mu\nu}{}^{\alpha'\beta'} = \delta^{\alpha'}_{(\alpha}\delta^{\beta'}_{\beta)}\delta^4(x,x')+ \nabla_{(\alpha}Z_{\beta)}{}^{\alpha'\beta'} \\ -\frac{1}{2}g_{\alpha\beta}\nabla_\mu Z^{\mu\alpha'\beta'}\,,
\end{multline} 
where $Z^{\alpha\alpha'\beta'}=\nabla_\beta  G^{\alpha\beta\alpha'\beta'}$. Now note that (i)~$\hat{\cal S}^{\alpha\beta}$ annihilates any linear gauge perturbation, meaning $\hat{\cal S}^{\alpha\beta}(\nabla_{(\alpha}\xi_{\beta)})=0$ for any $\xi_\alpha$ (even a distributional $\xi_\alpha$), and (ii) $\hat{\cal S}^{\alpha\beta}$ annihilates the trace part of any perturbation, meaning it annihilates any tensor proportional to $g_{\alpha\beta}$. Hence, the $Z_\alpha{}^{\alpha'\beta'}$ terms in Eq.~\eqref{eq:dGG} are annihilated by $\hat{\cal S}^{\alpha\beta}$, and Eq.~\eqref{eq:OTG} reduces to
\begin{align}\label{eq:OTG=Sdelta}
     \hat{\mathcal{O}} \hat{\mathcal{T}}^{\mu\nu} G_{\mu\nu\mu'\nu'} &= \hat{\mathcal{S}}_{\mu\nu}\bigl[\delta^\mu_{\mu'}\delta^\nu_{\nu'}\delta^4(x,x')\bigr]\,.
\end{align}
This acts on a test field at $x'$ as
\begin{equation}
    \langle f^{\mu'\nu'},\hat{\mathcal{O}} \hat{\mathcal{T}}^{\mu\nu} G_{\mu\nu\mu'\nu'}\rangle = \hat{\cal S}_{\mu\nu}f^{\mu\nu},
\end{equation}
exactly like Eq.~\eqref{eq:<phi,OGS>}. Thus, we have established $\psi^{(2)} = \hat{\mathcal{T}}^{\mu\nu} h^{(2)}_{\mu\nu}$ for $h^{(2)}_{\mu\nu}$ in any gauge. 

We have not explored how to establish the same result for Regge-Wheeler-Zerilli variables. In that case, the source operator does not annihilate traces; see Eq.~(189) of Ref.~\cite{Spiers:2023mor}, for example. We hence do not arrive at the analogue of Eq.~\eqref{eq:OTG=Sdelta}. Some other argument is then required, or simply a more specialized analysis of the trace term in Eq.~\eqref{eq:dGG}. 

An advantage of working with the scalar variables is that the equations are separable even in a Kerr background (in a basis of spin-weighted spheroidal harmonics). However, in Kerr this requires a Fourier transform, which introduces additional subtleties because $|\Delta r|^B$ depends on time for nonspherical orbits. For simplicity, we again restrict to a Schwarzschild background, in which case any of the scalar equations admit a spherical-harmonic decomposition. At the level of spherical modes, we can write
\begin{align}
    \hat{\mathcal{O}}_{\ell m} \psi_{\ell m}^{(2)} = S_{\ell m}^{(2)}.
\end{align}
The Green's-function representation of the solution, analogous to Eq.~\eqref{eq:hlm Greens function rep}, is
\begin{align}
    &\psi^{(2)}_{\ell m}  \nonumber\\
    &\ =8\pi\int G_{\ell m}(t-t',r,r')\hat{\mathcal{S}}^{a'b'}_{\ell m} \bigl( T_{a'b'}^{(2,\ell m)} - T^{Q,\ell m}_{a' b'}\bigr)\,dt' dr' \nonumber
    \\
    &\ \quad-\FP_{B=0}\int_{\rho<a}    \!\! \!|\Delta \tilde{r}|^B G_{\ell m}(t-t',r,r')\,  \hat{\mathcal{S}}^{a'b'}_{\ell m}\delta^2 G^{\ell m}_{a' b'} dt'dr' \nonumber
    \\
    &\ \quad+\int_{\rho>a}    G_{\ell m}(t-t',r,r')\, S_{\ell m}^{(2)}(t',r')\,dt' dr', \label{eq:psi2lm Greens}
\end{align}
where $G_{\ell m}$ is the Green's function for the mode-decomposed scalar equation. As in the four-dimensional case, one can show that the result is independent of the cut-off $a$ and that the placement of $|\Delta \tilde r|^B$ relative to $\hat{S}^{a'b'}_{\ell m}$ does not matter. The source term in the integrand, since it involves one or two derivatives of $\delta^2G^{\ell m}_{ab}$, is up to two orders more singular than the source in Eq.~\eqref{eq:ddGlm singularity}, with the form
 \begin{multline}\label{eq:Teuk mode source singularity}
    -\hat{\mathcal{S}}^{a'b'}_{\ell m}\delta^2 G^{\ell m}_{a' b'} \\= \frac{A_{\ell m}}{\Delta r'^4} +\frac{B_{\ell m}}{\Delta r'^3}+\frac{C_{\ell m}}{\Delta r'^2} +\frac{D_{\ell m}}{\Delta r'}+ \mathcal{O}(\log|\Delta r'|)\
\end{multline}
at points $r\neq r_p$\footnote{Note that the coefficients $A_{\ell m}$ and $B_{\ell m}$ here are different from the ones in Eq.~\eqref{eq:ddGlm singularity}.}. Note that all of the divergent terms arise from singular-singular terms; at the level of modes, singular-regular terms consist of jump discontinuities, delta functions, and their derivatives~\cite{Upton:2025bja}.

\subsection{Demonstration: static particle in flat spacetime}\label{sec:demonstration}

To demonstrate the methods laid out in the previous section, we consider the simplest scenario of the radial spin-weight $-2$ Teukolsky equation for a static particle at $r_p=r_0$ in flat spacetime:
\begin{align}
    \hspace{-2pt}\left[ r^2 \frac{d^2}{dr^2} - 2 r \frac{d}{dr} - (\ell-1)(\ell +2) \right] {}_{-2}\psi_{\ell m} = {}_{-2}T_{\ell m }, \label{eq:TeukolskyMaster}
\end{align}
where we assumed a static solution with no incoming waves. In this setting, the regular field vanishes ($h^{\R}_{\mu\nu} =0$), implying $\tilde T^{\mu\nu}_{(2)}=0=T^{\mu\nu}_{Q}$. The source in the second-order Einstein equation is then simply given by 
\begin{align}
    -\bigl\langle\varphi_{\mu\nu}, \widehat{\delta^2G}{}^{\mu\nu} \bigr\rangle  =- \bigl\langle \varphi_{\mu\nu}, {\rm{Pf}} \delta^2G^{\mu\nu}\bigr\rangle,
\end{align}
and the source $_{-2}T_{\ell m}$ has the form~\eqref{eq:Teuk mode source singularity}.

Requiring regularity as $r \rightarrow0$ and $r \rightarrow\infty$, we find the following Green's function for Eq.~\eqref{eq:TeukolskyMaster}:
\begin{multline}
    G_{\ell m}(r,r') = -\frac{1}{(2\ell +1)r'} \biggl[ \biggl( \frac{r'}{r}\biggr)^{\!\ell -1} \theta(r-r') \\+ \biggl( \frac{r}{r'}\biggr)^{\!\ell +2} \theta(r'-r)\biggr]\,.
\end{multline}
We can then find the solution using the one-dimensional equivalent of Eq.~\eqref{eq:psi2lm Greens}. Following the same strategy as in Eq.~\eqref{eq:FP example}, we add and subtract the most singular behaviour near the particle as given in ~\eqref{eq:Teuk mode source singularity}, yielding
\begin{widetext}
\begingroup \allowdisplaybreaks
\begin{align}
    _{-2}\psi_{\ell m}
    &= \FP_{B=0}\int_{|\Delta r'|<a} |\Delta \tilde{r}'|^B  G_{\ell m}(r,r')  _{-2}T_{\ell m}(r') dr' +\int_{|\Delta r'|>a}   G_{\ell m}(r,r')  _{-2}T_{\ell m}(r') dr'. \nonumber
    \\
    &= \int G_{\ell m}(r,r')  \left ( {}_{-2}T_{\ell m}(r') -\frac{A_{\ell m}}{\Delta r'^4} -\frac{B_{\ell m}}{\Delta r'^3}-\frac{C_{\ell m}}{\Delta r'^2} -\frac{D_{\ell m}}{\Delta r'}\right) dr' \nonumber
    \\
    &\quad +\FP_{B=0}\int_{|\Delta r'|<a} |\Delta \tilde{r}'|^B  G_{\ell m}(r,r')  \left( \frac{A_{\ell m}}{\Delta r'^4}+\frac{B_{\ell m}}{\Delta r'^3}+\frac{C_{\ell m}}{\Delta r'^2} +\frac{D_{\ell m}}{\Delta r'}\right) dr' \nonumber
    \\
    &\quad +\int_{|\Delta r'|>a}   G_{\ell m}(r,r')  \left( \frac{A_{\ell m}}{\Delta r'^4}+\frac{B_{\ell m}}{\Delta r'^3}+\frac{C_{\ell m}}{\Delta r'^2} +\frac{D_{\ell m}}{\Delta r' }\right) dr' \label{eq:93},
\end{align} 
where the integral in the second line above is now perfectly finite.

Formulas to evaluate Eq.~\eqref{eq:93} are given in Appendix~\ref{sec:FP integrals}. For example, using Eq.~\eqref{eq:Dterms}--\eqref{eq:Aterms} for $\ell=2$, we find that 
\begin{align}
    _{-2}\psi_{2 m} &=\int G_{2m}(r,r')  \left ( {}_{-2}T_{2m}(r') -\frac{A_{2m}}{\Delta r'^4} -\frac{B_{2m}}{\Delta r'^3}-\frac{C_{2m}}{\Delta r'^2} -\frac{D_{2m}}{\Delta r'}\right) dr' \nonumber
    \\
    &\quad +\frac{D_{2 m}}{60 r r_0^5}  \left[12 r^5 \log \left|1-\frac{r_0}{r}\right|+r r_0 \left(12 r^3+6 r^2 r_0+4 r r_0^2+3 r_0^3\right)-12 r_0^5 \log \left| \frac{r_0-r}{r_0} \right| \right] \nonumber
    \\
    &\quad -\frac{C_{2m}}{12 r_0^6} \left[12 r^4 \log \left|1-\frac{r_0}{r}\right|+r_0 \left(12 r^3+6 r^2 r_0+4 r r_0^2+3 r_0^3\right)\right]\nonumber
    \\
    &\quad -\frac{B_{2m}}{4 r_0^7} \left[\frac{r_0 \left(-12 r^4+6 r^3 r_0+2 r^2 r_0^2+r r_0^3+r_0^4\right)}{r-r_0}-12 r^4 \log \left| 1-\frac{r_0}{r}\right| \right] \nonumber
    \\
    &\quad -\frac{A_{2m}}{12 r_0^8} \left[84 r^4 \log \left|1-\frac{r_0}{r} \right|+\frac{r_0 \left(84 r^5-126 r^4 r_0+28 r^3 r_0^2+7 r^2 r_0^3+2 r r_0^4+3 r_0^5\right)}{(r-r_0)^2}\right].\label{eq:solution}
\end{align}
\endgroup
\end{widetext}
This is regular everywhere off the particle ($r \neq r_0$), and we can straightforwardly check that it is a solution to Eq.~\eqref{eq:TeukolskyMaster} for $r\neq r_0$. To verify that it satisfies Eq.~\eqref{eq:TeukolskyMaster} distributionally, we can once again integrate against a test function. 

In practical computations, one would typically carry out this calculation somewhat differently, by writing ${}_{-2}T_{\ell m} = -\hat{\cal S}^{ab}\delta^2G^{\ell m}_{ab}$ and integrating by parts to move $\hat{\cal S}^{ab}$ off the singular source and onto the Green's function~\cite{Leather:2026zhl}. That approach would reduce the number of singular powers of $\Delta r$ that need to be subtracted in the above procedure (to only the single power $1/\Delta r^2$ if the remaining integrals are assigned their principal value). We have presented the more brute-force approach for simplicity, to avoid the need to spell out any details of the operator $\hat{\cal S}^{ab}$.

\section{Conclusion}
\label{sec:discussion}

In this paper, we had two aims: to show how one can rigorously extend the traditional point-particle description, at the level of classical PDEs, beyond linear order in perturbation theory; and to put this description in a practical form that is useful for second-order black hole perturbation theory. 

The first aim was already partially achieved in our previous work~\cite{Upton:2021oxf}, where two of us showed how to formulate point-particle field equations at second perturbative order starting from the outputs of matched asymptotic expansions (a formalism where all fields are smooth and there is no singularity to regularize). However, we left the field equations in an unsatisfactory form that involved nontrivial distributions and still required analytical access to a puncture (the local singularity in the second-order metric perturbation), which should not be a requirement when solving directly for second-order retarded fields.

More importantly, there was a significant gap between the rigorous definition and any practical applications. Here, we have bridged this gap and provided a fully skeletonized version of the second-order field equation, Eq.~\eqref{eq:fullEFELorenz} with Eq.~\eqref{eq:ddGhat distribution}. This form is both more satisfactory and more practical, as it eliminates dependence on the puncture and involves only simple Dirac $\delta$ distributions. The upshot is a version of the field equations that is ready for practical applications in both analytical and numerical second-order self-force calculations. 

Our formulation of the field equations essentially comprises a small-scale cutoff around the particle's worldline, together with the skeletonized effective stress-energy tensor in Eq.~\eqref{eq:DetT}, and supplemented by the addition of two Dirac $\delta$-function sources. The first of these $\delta$ functions, $T^{\alpha\beta}_{\SS}$ in \eqref{eq:Tcount}, is a counterterm that cancels the divergence in the second-order solution when the cutoff scale is taken to zero. The second $\delta$ function, $T^{\alpha\beta}_{Q}$ in Eq.~\eqref{eq:TQ}, makes a finite contribution to the second-order metric perturbation and could not be derived from finiteness alone.

Although we assumed Lorenz gauge from the beginning, our results are more general than that. When deriving the distributional $\widehat{\delta^2G}{}^{\alpha\beta}$ in Eq.~\eqref{eq:ddGhat distribution}, we only relied on the leading singularity in $h^{\rm S(1)}_{\alpha\beta}$ being the same as in Lorenz gauge, along with the parity structure of the first subleading terms, described in Sec.~\ref{sec:local expansions}. One might suspect that the gauge of $h^{(2)}_{\alpha\beta}$ comes into play through the dependence on $h^{\S\S}_{\alpha\beta}$ in Eq.~\eqref{eq:d2GHat}, but that definition is invariant under linear transformations of $h^{\S\S}_{\alpha\beta}$ because $\deltaG^{\alpha\beta}[{\cal L}_\xi g]=0$. Thus, for most of the paper, the only assumption is that the first-order gauge is ``locally Lorenz'' in the sense of Ref.~\cite{Pound:2013faa} and has appropriate parity at order $\rho^0$. An exception is Sec.~\ref{sec:T2Conservation}. The derivation of the conservation of the effective stress-energy tensor from the Bianchi identity does rely on Eq.~\eqref{eq:QR near gamma}, where the second term in the expansion of $Q_\R^{\alpha \beta}[h^{\rm S (1)}]$ relies on $h^{\S(1)}_{\alpha\beta}$ being the same as in Lorenz gauge up to order $\mathcal{O}(\rho^0)$. More broadly, the original distributional promotion~\eqref{eq:d2GHat} can be used in any gauge compatible with the assumptions of matched asymptotic expansions (meaning $h^{(1)}_{\alpha\beta}\sim 1/\rho$), and we expect the only change to our results would be the concrete formulas for $T^{\alpha\beta}_{\S\S}$ and $T^{\alpha\beta}_{Q}$.

To demonstrate the utility and generality of our skeletonized field equations, we have explored several applications:
\begin{itemize}
    \item Our field equations allow one to obtain equations of motion from the conservation of the effective, skeletonized stress-energy tensor, rather than (or in addition to) deriving them from the field equations outside the body. We detailed this in Sec.~\ref{sec:T2Conservation}. 
    \item They can be used to show the validity of ``off the shelf'' regularization methods, such as the Hadamard regularization in Eq.~\eqref{eq:ddG = PfddG}, which we showed how to rigorously apply in Sec.~\ref{sec:Hadamard}.
    \item Finally, in Sec.~\ref{sec:applications} we described how our results can be practically implemented in black hole backgrounds. We showed how to obtain Green's-function representations of second-order solutions, how to apply our prescription at the level of spherical- or spheroidal-harmonic mode decompositions, and how to apply it to scalar equations rather than the Einstein equation, focusing particularly on the Teukolsky equation. We additionally included a concrete demonstration in a toy example.
\end{itemize}

This work invites many follow-ups. For example, our calculations could be extended to show the validity of dimensional regularization for the second-order field equation, which would be valuable in linking to the particle physics and EFT communities, where Hadamard regularization is less familiar. Our results should also enable simpler calculations of the second-order \emph{regular}, rather than retarded, field, and therefore the second-order self-force. This could be done using dimensional regularization, Hadamard regularization, or mode-sum regularization, for example, as alternatives to a puncture scheme. The last of these is the most standard method of computing the first-order regular field and self-force~\cite{Barack:2018yvs}, as it is typically easier to implement than a puncture scheme.

Another interesting avenue would be to examine what  distributional action would yield our distributional field equations when varied. This would further tighten the link to point-particle EFT methods that start from an action.

Finally, one could explore how our distributional definitions relate to the process of shrinking down a family of smooth material sources. Every distribution can be written as the limit of a family of smooth functions, which suggests that there should be a rigorous way of identifying our source as the limit of a family of smooth, extended sources. This might be best understood in the theory of nonlinear generalized functions on manifolds~\cite{Steinbauer:2006qi,Steinbauer:2008gn,Vickers:2012zz,Nigsch:2019tgl,Nigsch:2019uvt}, which would also enable extensions beyond second order. At third order and higher, $h^{(n)}_{\alpha\beta}\sim m^n/\rho^n$ is no longer locally integrable, and our trick of identifying nonlinear operators as linear operators on integrable functions would no longer suffice. The theory of nonlinear generalized functions could offer a way around that obstacle. For the reasons explained in Ref.~\cite{Damour:2001bu}, dimensional regularization could offer another viable way to extend our results to higher orders.

\begin{acknowledgments}

JM is supported by the Carlsberg Foundation, grant CF24-1716.
AP and SDU acknowledge the support of a Royal Society University Research Fellowship and the ERC Consolidator/UKRI Frontier Research Grant GWModels (as selected by the ERC and funded by UKRI under grant number EP/Y008251/1).

\end{acknowledgments}

\appendix

\section{Finite part integrals}
\label{sec:FP integrals}

In solving the scalar field equation in Sec.~\ref{sec:demonstration}, we have to deal with the following finite-part integrals:
\begin{align}
    \FP_{B=0}\int_{r_0-a}^{r_0+a}&  \left(  \frac{A_{\ell m}}{\Delta r'^4} +\frac{B_{\ell m}}{\Delta r'^3}+\frac{C_{\ell m}}{\Delta r'^2} +\frac{D_{\ell m}}{\Delta r'}\right) \nonumber
    \\
    &\,\times|\Delta \tilde{r}'|^B G_{\ell m} (r,r')dr'\,. 
\end{align}
Importantly, $G_{\ell m}(r,r')$ does not have compact support, unlike a test field. As a consequence, we pick up boundary terms when evaluating the finite part.

Specifically, we find that
\begingroup\allowdisplaybreaks
\begin{align}
    \hspace{-2pt}\FP_{B=0} &\int_{r_0-a}^{r_0+a} dr'  \frac{D_{\ell m}}{\Delta r'} |\Delta \tilde{r}'|^B  G_{\ell m}(r,r')\nonumber
    \\*
    &=\FP_{B=0} \int_{r_0-a}^{r_0+a} dr'  \frac{D_{\ell m}}{B} \partial_{r'} |\Delta \tilde{r}'|^B  G_{\ell m}(r,r')\nonumber
    \\
    &=-\int_{r_0-a}^{r_0+a} dr'D_{\ell m} \log\left({|\Delta r'|}/{a} \right) \partial_{r'} G_{\ell m}(r,r'), \label{eq:Dterms}
\end{align}
\begin{align}
    \hspace{-2pt}\FP_{B=0}& \int_{r_0-a}^{r_0+a} dr'  \frac{C_{\ell m}}{\Delta r'^2} |\Delta \tilde{r}'|^B  G_{\ell m}(r,r') \nonumber
    \\*
    &= \FP_{B=0} \int_{r_0-a}^{r_0+a} dr'  \frac{C_{\ell m}}{B(B-1)}\partial_{r'}^2 |\Delta \tilde{r}'|^B G_{\ell m}(r,r') \nonumber
    \\
    &=- \frac{C_{\ell m}}{a} \left[ G_{\ell m}(r,r_0+a)+G_{\ell m}(r,r_0-a) \right] \nonumber
    \\*
    &\quad - C_{\ell m} \int_{r_0-a}^{r_0+a}dr'\log \left( |\Delta r'| /a\right) \partial_{r'}^2 G_{\ell m}(r,r'), \label{eq:Cterms}
\end{align}
\begin{align}
    &\FP_{B=0} \int_{r_0-a}^{r_0+a} dr'  \frac{B_{\ell m}}{\Delta r'^3} |\Delta \tilde{r}'|^B  G_{\ell m}(r,r') \nonumber
    \\*
    &\ \ = \FP_{B=0} \int_{r_0-a}^{r_0+a} dr'  \frac{B_{\ell m}}{B(B-1)(B-2)}\partial_{r'}^3 |\Delta \tilde{r}'|^B G_{\ell m}(r,r') \nonumber
    \\
    &\ \ =- \frac{B_{\ell m}}{2a^2} \left[ G_{\ell m}(r,r_0+a)-G_{\ell m}(r,r_0-a) \right] \nonumber
    \\*
    &\ \ \quad - \frac{B_{\ell m}}{2a} \left[\partial_{r_0} G_{\ell m}(r,r_0+a)+\partial_{r_0}G_{\ell m}(r,r_0-a) \right] \nonumber
    \\*
    &\ \ \quad - \frac{1}{2}B_{\ell m}\int_{r_0-a}^{r_0+a}dr' \log \left(|\Delta r'|/a \right)  \partial_{r'}^3 G_{\ell m}(r,r'), \label{eq:Bterms}
\end{align}
and
\begin{align}
    &\FP_{B=0} \int_{r_0-a}^{r_0+a} dr'  \frac{A_{\ell m}}{\Delta r'^4} |\Delta \tilde{r}'|^B  G_{\ell m}(r,r') \nonumber
    \\*
    &\ \ = \FP_{B=0} \int_{r_0-a}^{r_0+a} dr'  \frac{A_{\ell m} \partial_{r'}^4 |\Delta \tilde{r}'|^B}{B(B-1)(B-2)(B-3)} G_{\ell m}(r,r') \nonumber
    \\*
    &\ \ =- \frac{A_{\ell m}}{3a^3} \left[ G_{\ell m}(r,r_0+a)+G_{\ell m}(r,r_0-a) \right] \nonumber
    \\*
    &\ \ \quad - \frac{A_{\ell m}}{6a^2} \left[\partial_{r_0} G_{\ell m}(r,r_0+a)-\partial_{r_0}G_{\ell m}(r,r_0-a) \right] \nonumber
    \\*
    &\ \ \quad - \frac{A_{\ell m}}{6a} \left[\partial^2_{r_0} G_{\ell m}(r,r_0+a)+\partial^2_{r_0}G_{\ell m}(r,r_0-a) \right] \nonumber
    \\*
    &\ \ \quad - \frac{1}{6}A_{\ell m}\int_{r_0-a}^{r_0+a}dr' \log \left(|\Delta r'|/a\right)  \partial_{r'}^4 G_{\ell m}(r,r'). \label{eq:Aterms}
\end{align}
We note that the integral in \eqref{eq:Dterms} can equivalently be written as a principle value integral:
\begin{align}
    \hspace{-2pt}\FP_{B=0} &\int_{r_0-a}^{r_0+a} dr'  \frac{D_{\ell m}}{\Delta r'} |\Delta \tilde{r}'|^B  G_{\ell m}(r,r')\nonumber
    \\*
    &= \text{P.V.} \int_{r_0-a}^{r_0+a} dr'  \frac{D_{\ell m}}{\Delta r'}   G_{\ell m}(r,r').
\end{align}
\endgroup

\bibliography{bibfile}
\end{document}